\documentclass{aa}  
\usepackage{upgreek}
\usepackage{amsmath}
\usepackage{float}
\usepackage{xspace}

\usepackage{pdflscape}
\usepackage{graphicx}
\usepackage{txfonts}

\usepackage[normalem]{ulem}
\usepackage{lipsum}

\usepackage{multirow}

\usepackage{hyperref}
\hypersetup{
    colorlinks=true,
    citecolor=blue,
    linkcolor=blue,
    filecolor=blue,
    urlcolor=blue
    }
    
\begin{document}

\renewcommand{\thempfootnote}{\arabic{mpfootnote}}

\newcommand{\formosa}{\texttt{ForMoSA}}

\newcommand{\Rjup}{\ensuremath{\mathrm{R_{Jup}}}\xspace}
\newcommand{\MJup}{\ensuremath{\mathrm{M_{Jup}}}\xspace}
\newcommand{\Mjup}{\MJup\xspace} 
\newcommand{\Msun}{\ensuremath{\mathrm{M_{\odot}}}}
\newcommand{\kms}{km/s\xspace}
\newcommand{\um}{$\upmu$m\xspace}
\newcommand{\Mdot}{\ensuremath{\dot{\textrm{M}}}}
\newcommand{\PaBeta}{\ensuremath{\mathsf{Pa}_{\upbeta}}\xspace}
\newcommand{\Halpha}{\ensuremath{\mathsf{H}_{\upalpha}}\xspace}
\newcommand{\Hbeta}{\ensuremath{\mathsf{H}_{\upbeta}}\xspace}
\newcommand{\BrGamma}{\ensuremath{\mathsf{Br}_{\upgamma}}\xspace}

\newcommand{\KI}{\ensuremath{\mathrm{K\,\textsc{i}}}\xspace}
\newcommand{\HI}{\ensuremath{\mathrm{H\,\textsc{i}}}\xspace}

\newcommand{\Pb}{\PaBeta} 
\newcommand{\Ha}{\Halpha} 
\newcommand{\Brg}{\BrGamma} 

\newcommand{\gqlupa}{GQ~Lup~A}
\newcommand{\gqlupb}{GQ~Lup~b}
\newcommand{\gsc}{GSC~06214-00210~b}
\newcommand{\shortgqlupa}{GQ~Lup~A}
\newcommand{\shortgqlupb}{GQ~Lup~b}
\newcommand{\shortgscb}{GSC06214b}
\newcommand{\shortgsca}{GSC06214A}

\newcommand{\Rin}{$\mathrm{R}_{\mathrm{in}}$}
\newcommand{\width}{$\textrm{width}$}
\newcommand{\MdotMagAcc}{\ensuremath{\dot{\textrm{M}}}}
\newcommand{\Tmax}{$\textrm{T}_{\textrm{max}}$}
\newcommand{\inclination}{$\textrm{inclination}$}
\newcommand{\mass}{$\textrm{M}$}
\newcommand{\Rp}{$\textrm{R}_{\textrm{p}}$}
\newcommand{\Teff}{\ensuremath{\textrm{T}_{\textrm{eff}}}\xspace}

\newcommand{\vz}{\ensuremath{\textrm{v}_{0}}}
\newcommand{\nz}{$\textrm{n}_{0}$}
\newcommand{\Seff}{$\textrm{S}_{\textrm{eff}}$}
\newcommand{\ffill}{\ensuremath{\textrm{f}_{\textrm{fill}}}\xspace}

\newcommand{\todo}[1]{\textbf{TODO #1}}
\newcommand{\toadd}[1]{\textbf{TOADD #1}}
\newcommand{\okish}[1]{\textit{\textbf{#1}}}

\authorrunning{D. Demars et al.}
\titlerunning{Emission line variability of GQ Lup b and GSC 06214-00210 b}

   \title{Emission line variability of young 10--30~\MJup\ companions\thanks{Based on observations collected at the European Organisation for Astronomical Research in the Southern Hemisphere under ESO program 099.C-0760.}}

   \subtitle{I. The case of GQ Lup b and GSC 06214-00210 b}

   \author{D. Demars\inst{1},
   M. Bonnefoy\inst{1},
   C. Dougados\inst{1},
   Y. Aoyama\inst{2, 3, 4},
   T. Thanathibodee\inst{5, 6},
   G.-D. Marleau\inst{7, 8, 9, 10},
   P. Tremblin\inst{11},
   P. Delorme\inst{1},
   P. Palma-Bifani\inst{12},
   S. Petrus\inst{13, 14},
   B. P. Bowler\inst{15},
   G. Chauvin\inst{12, 1},
   A.-M. Lagrange\inst{16, 1}}

   \institute{Univ. Grenoble Alpes, CNRS, IPAG, F-38000 Grenoble, France \\
              \email{dorian.demars@univ-grenoble-alpes.fr}
    \and
    Institute for Advanced Study, Tsinghua University, Beijing 100084, People’s Republic of China
    \and
    Department of Astronomy, Tsinghua University, Beijing 100084, People’s Republic of China
    \and
    Department of Earth and Planetary Science, The University of Tokyo, 7-3-1 Hongo, Bunkyo-ku, Tokyo 113-0033, Japan
    \and
    Department of Astronomy, University of Michigan, 1085 S. University Ave., Ann Arbor, MI 48109, USA
    \and
    Institute for Astrophysical Research and Department of Astronomy, Boston University, 725 Commonwealth Ave., Boston, MA 02215, USA
    \and
    Fakult\"{a}t f\"{u}r Physik, Universit\"{a}t Duisburg-Essen, Lotharstraße 1, 47057 Duisburg, Germany
    \and
    Institut f\"ur Astronomie und Astrophysik, Universit\"at T\"ubingen, Auf der Morgenstelle 10, D-72076 T\"ubingen, Germany
    \and
    Physikalisches Institut, Universit\"{a}t Bern, Gesellschaftsstr.~6, CH-3012 Bern, Switzerland
    \and
    Max-Planck-Institut f\"{u}r Astronomie, K\"{o}nigstuhl 17, 69117 Heidelberg, Germany
    \and
    Maison de la Simulation, CEA, CNRS, Univ. Paris-Sud, UVSQ, Université Paris-Saclay, 91191 Gif-sur-Yvette, France
    \and
    Université Côte d'Azur, OCA, Lagrange CNRS, 06304 Nice, France
    \and
    Instituto de F\'{i}sica y Astronom\'{i}a, Facultad de Ciencias, Universidad de Valpara\'{i}so, Av. Gran Breta\~{n}a 1111, Valpara\'{i}so, Chile
    \and
    Núcleo Milenio Formación Planetaria - NPF, Universidad de Valparaíso, Av. Gran Bretaña 1111, Valparaíso, Chile
    \and
    Department of Astronomy, The University of Texas at Austin, 2515 Speedway, Stop C1400, Austin, TX 78712, USA
    \and
    LESIA, Observatoire de Paris, PSL Research University, CNRS, Sorbonne Universités, UPMC Univ. Paris 06, Univ. Paris Diderot, Sorbonne Paris Cité, 5 place Jules Janssen, 92195 Meudon, France}

   \date{Received \dots; accepted \dots}

  \abstract
   {Emission lines indicative of active accretion have been seen on a handful of low-mass companions (M<30~\MJup) to stars. Line variability is ubiquitous on stellar accretors but has never been characterized in detail on low-mass companions and can give insights on the accretion mechanism at play.}
   {We aim to characterize the short- to long-term \HI\ Paschen $\upbeta$ emission line variability of two 10 to 30~\MJup\ companions on wide orbits -- \object{\gqlupb} and \object{\gsc} --  to clarify the accretion mechanisms at play.}
   {We used J-band observations at R=1800--2360, obtained with VLT/SINFONI in 2017, to record time series investigating hours-to-week variability of the \HI\ Paschen $\upbeta$ emission line (1.282~\um). Contrary to \Halpha, it should be less affected by chromospheric activity contamination. The photospheric emission is analyzed at each epoch and removed with the \texttt{FORMOSA} forward modeling tool, using new grids of ATMO models exploring different C/O and [M/H] values. The time series of line profiles and intensities are compared to those of more massive accretors and to predictions from the latest magnetospheric accretion and shock models. To complement, we also re-investigate archival spectroscopic observations at near-infrared wavelengths of each target to increase the time span up to a decade and build a more comprehensive understanding of the variability processes at play.}
   {For \gqlupb, we find line variability on months and decade timescales, while it is within noise level at shorter timescales. For \gsc, we find line variability on timescales of tens of minutes up to a decade. The line profiles of \gsc\ are partly resolved in at least one epoch. Both objects show \HI\ Paschen $\upbeta$ flux variability that is moderate ($<50$~\%) at timescales below their rotation period, and more important at longer timescales (up to $\sim1000$~\% on decade timescales), a behavior similar to that of classical T Tauri stars. The line profiles of \gqlupb\ are blue-shifted and can only be reproduced by magnetospheric accretion models, while \gsc's are fairly well reproduced by both magnetospheric accretion and shock models, except for one epoch for which the shock model is highly favored. The companions have C/O values broadly consistent with solar values.}
   {While magnetospheric accretion is favored for \gqlupb, higher resolution ($\textrm{R}>10\,000$) observations are required to disentangle the two (non-exclusive) emitting mechanisms. The similar variability behavior observed in these low mass companions and in classical T Tauri stars may support similar accretion mechanisms. The high amplitude of variability at months and longer timescales found in both objects could be key to explain the low yield of \Halpha\ imaging campaigns.}

   \keywords{Planets and satellites: formation, individual: GQ Lup b, GSC 06214-00210 b - Accretion, accretion disks}

   \maketitle
%

\section{Introduction}

Gas accretion on forming giant planets and brown-dwarf companions is a critical step determining their early physical evolution and internal structure \citep[e.g.,][]{2007ApJ...655..541M, 2009ApJ...702L..27B, 2018MNRAS.477.4817C} and the dissipation of their angular momentum \citep[e.g.,][]{2018NatAs...2..138B, 2018AJ....155..178B}. It may also drive disk--planet migration \citep[e.g.,][]{2016MNRAS.462.4130P} and the formation and water content of exo-moons \citep[e.g.,][]{2015ApJ...806..181H} which are possible habitats for life.

The theoretical understanding of the accretion processes is still in its infancy. Companions should accrete part of their mass from circumplanetary disks \citep[hereafter CPDs,][]{1982PThPh..67.1053M}  forming around them. Boundary-layer accretion is  proposed to develop if the CPD hits the planet surface \citep[e.g.,][and ref. therein]{2021ApJ...921...54D, 2022arXiv221202311F} producing a shock within a thin boundary at the disk-planet interface.  The most recent radiative-hydrodynamic (RHD) simulations of planets embedded in circumplanetary-disks predict that part of the mass should fall onto the planet embryo from the circumstellar disk via high-altitude flows and producing an accretion shock at the CPD surface or directly at the planet surface (e.g., \cite{2014ApJ...782...65S}, or Fig.~1 in \cite{szulagyi_hydrogen_2020} for a representation). 

The picture of accretion may in fact not be fully captured by the RHD approach. The strength of the dipole component of Jupiter’s current magnetic field is only about 4 Gauss \citep{1974Sci...183..305S}, but the interior of young exoplanets and low-mass brown-dwarfs are predicted to be strongly convective and ionized, and so be able to fuel $\sim$~kG magnetic fields \citep{reiners_magnetic_2010} sufficient to affect the gas flow \citep{2018A&A...619A.165C} and truncate the CPD \citep{2003A&A...411..623F}. It thus remains to be known to what extent the magnetospheric accretion framework applies in the planetary-mass regime.

The heated gas, either at the accretion shock or within the funnels, produces a set of emission lines from the UV to the NIR whose intensity and spectral profiles start to be predicted by accretion models \citep[e.g.,][]{thanathibodee_magnetospheric_2019, szulagyi_hydrogen_2020, aoyama_spectral_2021, 2022RNAAS...6..262M}. These lines have now been found on a handful of young  directly-imaged  young  companions with mass below $\sim$30\MJup\ \citep[TWA 5 B, GQ Lup B, CT Cha B, USco CTIO 108 B, DH Tau B, GSC 06214-00210 B, SR 12 C, PDS 70 b and c, 2M0249c, Delorme 1(AB)b, HIP 77900B, HIP 78530B,  USco 161031.9-16191305B, YSES 1b;][]{2000A&A...360L..39N, 2007A&A...463..309S,2008A&A...491..311S,2008ApJ...673L.185B, Bonnefoy14, bowler_disk_2011, sm18,sm19, 2018ApJ...863L...8W, 2014ApJ...783L..17Z, 2019NatAs...3..749H, 2021A&A...645A..17C, petrus_new_2020, 2021Natur.595..370Z}  and are estimated to be frequent \citep[$\sim$50\% frequency,][]{2017AJ....154..165B} below 15~Myr. These companions are found in systems spanning a wide range of ages (1--40~Myr)  and with diverse architectures (within the cavity of circumstellar disks, in hierarchical systems, or orbiting single or binary stars of various masses).

All the line detections have also been fortuitous and follow up-studies are starting to understand their origin and constrain the accretion models. \cite{2014ApJ...783L..17Z} and \cite{2021AJ....161..244Z} recorded measurements of the shock emission of four companions in the UV (\gqlupb, \gsc, DH~Tau~b and PDS~70~b, all within 10--30~\Mjup and $<$15~Myr). They could measure accretion rates in the range $\mathrm{10^{-9}-10^{-11}}$~$\Msun\,\mathrm{yr^{-1}}$.
They also argue that a significant fraction of the energy released at the accretion shock is reprocessed at \Halpha. Line ratios measurements on Delorme~1~(AB)b \citep[$\sim$14~\Mjup, $\sim$40~Myr,][]{delorme_direct-imaging_2013} at near-infrared wavelengths \citep{2022ApJ...935L..18B,betti22err} were found to be compatible with non-LTE accretion-shock models for planets \citep{aoyama_theoretical_2018}. High-resolution observations of the Balmer lines of Delorme~1~(AB)b in the UV show line profiles suggestive of magnetospheric accretion operating on that 40~Myr old companion \citep{2023A&A...669L..12R}. Constraints on the \Ha/\Hbeta line-ratio of PDS~70~b indicate significant extinction in the line-of-sight due to surrounding dust or a CPD \citep{2020AJ....159..222H}.

Variability of emission lines is observed across a wide stellar mass spectrum \citep[for a review, see][]{2022arXiv220311257F}, and traces episodic variability in the accretion and ejection over the course of the stars evolution. Non-steady accretion is predicted to occur on low-mass companions and has a significant impact on their early physical evolution \citep{2009ApJ...702L..27B, 2018MNRAS.477.4817C} and our ability to detect them \citep{2020ApJ...895...48B}. Global hydrodynamic and magneto-hydrodynamic simulations of protoplanets embedded in a circumstellar disk indicate modulations on decade-long timescales \cite[see][]{2013ApJ...779...59G, szulagyi_hydrogen_2020}. Instabilities or dead zones developing through the CPD \citep{2012ApJ...749L..37L, 2016ApJ...832..193Z}  should modulate the accretion flow at Keplerian timescales (months to years). If magnetospheric accretion operates,  hot spots and funnel flows  should produce variability of the emission lines along the object rotation period alike on stars \citep[e.g.,][]{2015A&A...580A..82S}.  Variable extinction in the line of sight should modulate the line emission further \citep{szulagyi_hydrogen_2020, marleau_accreting_2022}.

The study of the emission line variability timescales and amplitude therefore offers to probe both the accretion mechanism and the effect of the local environment of young companions. Line variability has been detected at the $\sim$year timescale on \gqlupb\ and \gsc\ (see below). \cite{wolff_upper_2017} found the \PaBeta\ line of DH~Tau~b \citep[a $\sim$1Myr old $\sim15$\MJup\ companion,][]{itoh_young_2005} to disappear within $\sim5$weeks. \cite{eriksson_strong_2020} find the \Halpha\ line of Delorme~1(AB)b \citep[a 30--40 Myr old, $\sim 14$\MJup\ companion][]{delorme_direct-imaging_2013} to vary over a few hours. \cite{2021AJ....161..244Z} report several measurements of the \Halpha\ line of PDS~70~b over days to month timescales and find an upper limit of 30\% on the variability.

We focus in this paper on the spectroscopic monitoring of the \PaBeta\ emission line of two low-mass companions -- \gqlupb\ and \gsc\ -- with previously reported line variability over the course of a semester \citep{lavigne_near-infrared_2009}. We combine these observations with archival data to probe line variability from minutes to decade timescales. The \PaBeta\ line is  well adapted to a monitoring with standard adaptive-optics fed instruments from the ground and is more intense above the continuum than the \BrGamma\ line in young accreting late-type objects \citep{2004A&A...424..603N, 2022ApJ...935L..18B}. Moreover, it is known to be less affected by chromospheric activity than \Halpha\ in the stellar and sub-stellar mass regimes \citep{2004A&A...424..603N, 2011A&A...534A..32A, 2012A&A...548A..56R, 2013A&A...551A.107M}.  

We present a compendium of the presently known properties of the two objects in Section \ref{sec:section1}. The observations of \gqlupb\ and \gsc\ are detailed in Section \ref{sec:obsdata} followed by the data extraction strategy in Section \ref{sec:extraction}. The photospheric parameters, and emission line variability (timescale, intensity, profiles) are presented in Section \ref{sec:results} and compared to models. We analyze how the \PaBeta\ line behavior relates to other accretors, contextualize possible line profile variations,  and discuss the implication of line variability for the detectability of protoplanets in Section \ref{sec:discussion}. Results are summarized in Section \ref{sec:conclusions}.

\section{Information on targets}
\label{sec:section1}
\subsection{GQ Lup system}
\label{subsec:GQLupb}
\shortgqlupb\ is a 10 to 40~\MJup\ companion resolved by \cite{2005A&A...435L..13N} at $\sim$0.7" \citep[$\sim$108~au at $154.10\pm0.69$~pc;][]{2022yCat.1355....0G} from a 2-5~Myr old K7Ve-type T Tauri star \citep[$\textrm{M}_{\star}=1.02 \pm 0.05$~\Msun;][]{2017ApJ...835...17M, 2022yCat.1355....0G} member of the Lupus I cloud \citep{1977ApJ...214..747H}.
The star is surrounded by a circumstellar disk inclined at 60$^{\circ}$ showing spiral structures \citep[e.g.,][]{van_holstein_survey_2021} in scattered light up to 0.84" (129~au), while the disk shows smaller extension at millimeter and sub-millimeter wavelengths both in continuum and $^{12}$CO \citep[20--46~au;][]{2017ApJ...835...17M, 2017ApJ...836..223W, 2020ApJ...895L..46L}.
\cite{2020ApJ...895L..46L} report a  gap at 10~au in the dust component of the disk at millimeter wavelengths which might be carved by a planet at this separation. It is unclear whether this possible planet could explain the significant radial-velocity variation of GQ Lup A noticed by \cite{2012MNRAS.425.2948D} on a 2 years time span. The star has a reported spin axis inclination of $\sim$30$^{\circ}$ misaligned with the disk and is experiencing dominantly pole-ward magnetospheric accretion that produces a set of emission lines, including \PaBeta\ \citep{2007A&A...468.1039B, 2012MNRAS.425.2948D, 2017A&A...602A..33F, 2007A&A...463..309S}. 

\shortgqlupb\ is also known to display a bright \Halpha\ emission line found at four epochs spanning two decades: on April 10, 1999 HST/WFPC2 wide-field imaging data  \citep[F606W filter; $\uplambda_{cen}=605.4$, $\textrm{FWHM}=220.5$~nm;][]{2007ApJ...654L.151M}, in HST/WFC3-UVIS2 data \citep{2014ApJ...783L..17Z} acquired on February 25, 2012 with a narrower filter (F656N, $\uplambda_{\textrm{cen}}=656.2~nm, \textrm{FWHM}=187.1$~nm); on ground-based Magelan/MagAO imaging \citep{2017ApJ...836..223W} with a custom \Halpha\ filter on April 16, 2015 ($\uplambda_{\textrm{cen}}=656.0~nm, \textrm{FWHM}=6.3$~nm), and with VLT/MUSE medium-resolution (R=2516) optical spectroscopy on April 19, 2019 \citep{2021AJ....162..286S}. \cite{2017ApJ...836..223W} argues that the line could be showing variability, but the different instrumental setup prevents from conducting a robust homogeneous comparison of the reported \Halpha\ fluxes. The MUSE spectrum shows the Ca~II triplet in emission ($\uplambda$8498, 8542, 8662~\AA) but no \Hbeta\ line.

The companion displays a \PaBeta\ emission line first seen on September 18, 2006 (EW=$-3.83\pm 0.12$~\AA) from data acquired with VLT/SINFONI \citep{2007A&A...463..309S} at R=2360. The line is also detected in Gemini/NIFS data (R$\sim$5000) collected on May 30, 2017 but at a significantly weaker flux level (EW=$-0.46\pm0.08$\AA),  supporting the observed variability at \Halpha. No \BrGamma\ emission line is reported in K-band spectra obtained at four epochs \citep{2005A&A...435L..13N, mcelwain_first_2007, 2007A&A...463..309S, lavigne_near-infrared_2009}.

The set of available near-infrared spectra revealed an inconsistency in the spectral slope of the SINFONI spectrum of \cite{2007A&A...463..309S} that we re-investigate in Section \ref{sec:extraction}. The compendium of 0.6--5~\um spectro-photometry on the companion allowed \cite{2021AJ....162..286S} to derive \Teff of 2700~K consistent with the optical spectral type of M9 \citep[e.g,][]{2003ApJ...593.1093L} and the luminosity-based ``hot-start'' mass of 30~\MJup. The analysis also confirms a significant in-line extinction ($\textrm{Av}=2.3$~mag) and an excess emission long-wards of 2.5~\um that could be caused by infalling material from the circumplanetary disk or a CPD. ALMA observations have failed to detect this CPD thus far \citep{2017ApJ...835...17M, 2017ApJ...836..223W} and no intrinsic linear polarization is detected at the location of the companion \citep{van_holstein_survey_2021}.
High-resolution spectroscopy (VLT/CRIRES, R$\sim$100~000) of \shortgqlupb\ centered on the $^{12}$CO overtone at K-band allowed to measure a  projected rotational velocity \citep[$5.3^{+0.9}_{-1.0}$~\kms;][]{2016A&A...593A..74S}, in the lower range of values measured on young 10-50~\MJup\ objects \citep{2006MNRAS.372.1879K, 2018NatAs...2..138B,2020ApJ...905...37B} which sets an upper limit on the rotation period of 82h.

\shortgqlupb's orbital motion has been monitored since its original discovery \citep{2014MNRAS.444.2280G, 2016A&A...593A..74S} and now covers $\sim1$~\% of its estimated orbit. Tentative constraints of the mutual inclination with the circumstellar disk indicate the companion might be on a polar orbit \citep{2021AJ....162..286S} with a semi-major axis of $\textrm{a} = 117^{+24}_{-23}$~au and low eccentricity possibly crossing the circumstellar disk. 

A second accreting low-mass stellar companion (2MASS 15491331-3539118; projected separation $\sim2400$~au; $M\sim0.15$~\Msun, GQ~Lup~C) was recently found to be bound to the star. GQ~Lup~c bears its own detached disk roughly aligned with the one of the primary star and with the large-scale dust filament of Lupus I containing GQ~Lup \citep{2020A&A...635L...1A, 2020A&A...635L..11L}. 

In summary, this system shows companions of different masses and provide a unique mean to better understand the early formation and dynamical evolution of planetary systems.

\subsection{GSC 06214-00210 system}
\label{subsec:GSC06214b}

\gsc\ (hereafter \shortgscb) is a $14.5 \pm 2$~\MJup\ co-moving companion \citep{2008ApJ...679..762K, 2011ApJ...726..113I, 2019AJ....157...71P} resolved on a wide orbit  \citep[2.2" separation, corresponding to a projected physical separation of 240~au at $108.8\pm0.3$~pc;][]{2022yCat.1355....0G} from the weak-line T Tauri star GSC~06214-00210 \citep[K$5\pm1$; $\textrm{M}_{\star}= 0.80\pm0.11$~\Msun;][]{2014ApJ...784...65B, 2019AJ....157...71P}. 

The star is reported to be a member of the Upper Scorpius sub-group of the Scorpius-Centaurus (Sco-Cen) OB association \citep{1998A&A...333..619P}, with a median age of $\sim10$~Myr \citep{2012ApJ...746..154P, 2016A&A...593A..99F}. We notice that the kinematic analysis tool BANYAN $\Sigma$ \citep{2018ApJ...856...23G} now suggests a possible alternative membership (58.8\% probability) to the older Upper-Scorpius Centaurus (UCL) sub-group of Sco-Cen \citep[16 Myr;][]{2012ApJ...746..154P} based on the GAIA-DR3 astrometry. This possible membership to UCL is better consistent with the revised isochronal age of $24^{+7}_{-5}$~Myr reported in \cite{2019AJ....157...71P} and based on the \cite{2015A&A...577A..42B} evolutionary tracks and their $16.9^{+2.9}_{-1.9}$~Myr age inferred using the magnetized evolutionary models of \cite{2016A&A...593A..99F} and a prior on the age of Sco-Cen. We adopt this later value in our study. 

The star is known to have no reported excess emission, a weak \Halpha\ emission line, and no \PaBeta\ emission \citep{1998A&A...333..619P, bowler_disk_2011}. Conversely, medium-resolution 1.17–1.83~\um\ spectroscopy (R=3800) of the system with OSIRIS at Keck in 2010 \citep{bowler_disk_2011} revealed a bright \PaBeta\ emission line (EW$ =-11.4\pm0.3$~\AA) on \shortgscb. HST/WFC3 photometry also reveals the companion to emit at \Halpha\ \citep{2014ApJ...783L..17Z}. The companion has a continuum excess emission noticeable at K- and L'- bands \citep[1.8-4.2~\um;][]{bowler_disk_2011, van_holstein_survey_2021} which indicates it likely bears a CPD. The CPD has remained elusive at sub-millimeter wavelengths \citep{2015ApJ...805L..17B, 2017AJ....154..234W} indicative of a compact detached disk. A likely intrinsic polarization signal \citep{van_holstein_survey_2021} was however detected with VLT/SPHERE. \cite{van_holstein_survey_2021} argue that the CPD must have a inclination of $\sim48^{\circ}$ to explain the measured degree of polarization. They used the possible constraints on the CPD inclination and the measured rotational velocity of $6.1^{+4.9}_{-3.8}$~\kms\ of \shortgscb\ \citep{2018NatAs...2..138B} to narrow down the companion rotation period to 22–77~h. 

An additional 0.9--2.5~\um\ spectrum of \shortgscb\ was obtained with GNIRS on June 23, 2011 by \cite{lachapelle_characterization_2015} with a resolution lower than OSIRIS (R$\sim$1000). It shows a weaker \PaBeta\ emission (EW$ =-4.4\pm0.5$~\AA) than in \cite{bowler_disk_2011} and a \BrGamma\ emission line (EW$=-0.24\pm0.05$~\AA), making \shortgscb\ the third companion with mass below 20~\MJup\ to display this line \citep{2021Natur.595..370Z, 2022ApJ...935L..18B}.
K-band (1.965–2.381~\um) OSIRIS spectra obtained on June 26, 2012 at R=3800 do not show the reported \BrGamma\ line. \cite{2018NatAs...2..138B} do not mention the detection of this line in their R=25000 Keck/NIRSPEC spectrum (2.03-2.38~\um) of the target obtained on June 3, 2015 either. But this line is re-detected in archival VLT/SINFONI spectra (R$\sim$5000) obtained on May 28, 2014 and July 07, 2014 (Palma-Bifani et al., in preparation) further hinting for significant year-to-year variability of accretion lines on that companion.  All sets of near-infrared spectra allow to classify this companion as a M9-M9.5 low-gravity object \citep{lavigne_near-infrared_2009, 2014ApJ...784...65B}. 

\shortgscb\ offers to probe the accretion processes at the Deuterium-burning boundary in a more evolved system, thus complementing \shortgqlupb's analysis.

\section{Observations and data reduction} 
\label{sec:obsdata}

\longtab{

\begin{longtable}{llcccccccccc}
\caption{Log of SINFONI observations. Dates are given in the YYYY-MM-DD format. DIT and NDIT and correspond to the Detector Integration Time per frame, the number of individual frames per exposure. $\mathsf{\uptau_{0}}$ corresponds to the coherence time. Both seeing and coherence time were gathered from the ESO/ASM tool at corresponding UT times, while the airmass was computed as the mean value between the starting and ending values in the files headers. The J-mag is our computed J-2MASS magnitude. $\Delta\uplambda$ corresponds to the applied wavelength calibration, as described in Section~\ref{sec:cube_reconstruction}, i.e., the value found for the telluric standard of that sequence.}\\ 

\hline\hline
Target & Date & UT start & UT end & Setup & $\mathrm{DIT\times NDIT}$ & seeing & airmass & $\mathsf{\uptau_{0}}$ & J-mag & $\Delta\uplambda$ \\
&  & (h:m:s) & (h:m:s) & mas/px & (s) & (") & & (ms) & & \AA\\
\hline
\endfirsthead
\shortgqlupb & 2017-04-25 & 05:21:16 & 05:26:16 & 25 & 300$\mathrm{\times1}$ & 0.62 & 1.04 & - & 14.59 & $1.32\pm0.21$\\
\shortgqlupb & 2017-04-25 & 05:26:40 & 05:31:40 & 25 & 300$\mathrm{\times1}$ & 0.56 & 1.04 & - & 14.43 & $1.32\pm0.21$\\
\shortgqlupb & 2017-04-25 & 05:37:30 & 05:42:30 & 25 & 300$\mathrm{\times1}$ & 0.56 & 1.03 & - & 14.35 & $1.32\pm0.21$\\
\shortgqlupb & 2017-04-25 & 05:42:55 & 05:47:55 & 25 & 300$\mathrm{\times1}$ & 0.43 & 1.03 & - & 14.10 & $1.32\pm0.21$\\
\shortgqlupa & 2017-04-25 & 05:54:10 & 05:54:25 & 25 & 15$\mathrm{\times1}$ & 0.64 & 1.02 & - & - & $1.32\pm0.21$\\
Hip081208 & 2017-04-25 & 06:05:03 & 06:05:06 & 25 & 3$\mathrm{\times1}$ & 0.65 & 1.04 & - & - & $1.32\pm0.21$\\
\hline
\shortgqlupb & 2017-04-30 & 05:34:35 & 05:39:35 & 25 & 300$\mathrm{\times1}$ & 0.45 & 1.02 & - & 13.95 & $0.95\pm0.21$\\
\shortgqlupb & 2017-04-30 & 05:56:16 & 06:01:16 & 25 & 300$\mathrm{\times1}$ & 0.44 & 1.02 & - & 14.24 & $0.95\pm0.21$\\
\shortgqlupb & 2017-04-30 & 05:50:51 & 05:55:51 & 25 & 300$\mathrm{\times1}$ & 0.43 & 1.02 & - & 14.27 & $0.95\pm0.21$\\
\shortgqlupb & 2017-04-30 & 05:40:00 & 05:45:00 & 25 & 300$\mathrm{\times1}$ & 0.40 & 1.02 & - & 13.96 & $0.95\pm0.21$\\
\shortgqlupa & 2017-04-30 & 06:07:36 & 06:07:51 & 25 & 15$\mathrm{\times1}$ & 0.58 & 1.02 & - & - & $0.95\pm0.21$\\
Hip079410 & 2017-04-30 & 06:17:30 & 06:17:34 & 25 & 4$\mathrm{\times1}$ & 0.53 & 1.00 & - & - & $0.95\pm0.21$\\
\hline
\shortgqlupb & 2017-06-01 & 06:29:29 & 06:34:29 & 25 & 300$\mathrm{\times1}$ & 0.60 & 1.24 & 15.51 & 14.14 & $1.34\pm0.18$\\
\shortgqlupb & 2017-06-01 & 06:34:55 & 06:39:55 & 25 & 300$\mathrm{\times1}$ & 0.57 & 1.25 & 13.92 & 14.26 & $1.34\pm0.18$\\
\shortgqlupb & 2017-06-01 & 06:45:44 & 06:50:44 & 25 & 300$\mathrm{\times1}$ & 0.69 & 1.29 & 9.70 & 14.53 & $1.34\pm0.18$\\
\shortgqlupb & 2017-06-01 & 06:51:09 & 06:56:09 & 25 & 300$\mathrm{\times1}$ & 0.70 & 1.31 & 10.15 & 14.50 & $1.34\pm0.18$\\
\shortgqlupa & 2017-06-01 & 07:02:25 & 07:02:40 & 25 & 15$\mathrm{\times1}$ & 0.67 & 1.35 & 9.85 & - & $1.34\pm0.18$\\
Hip082670 & 2017-06-01 & 07:12:57 & 07:13:17 & 25 & 20$\mathrm{\times1}$ & 0.56 & 1.26 & 11.90 & - & $1.34\pm0.18$\\
\hline
\shortgqlupb & 2006-09-18 & 23:22:53 & 23:27:53 & 25 & 300$\mathrm{\times1}$ & 0.87 & 1.24 & 1.96 & 13.01 & $1.30\pm0.16$\\
\shortgqlupb & 2006-09-18 & 23:39:49 & 23:44:49 & 25 & 300$\mathrm{\times1}$ & 0.93 & 1.30 & 1.74 & 12.79 & $1.30\pm0.16$\\
\shortgqlupb & 2006-09-18 & 23:45:21 & 23:50:21 & 25 & 300$\mathrm{\times1}$ & 0.86 & 1.32 & 1.95 & 13.06 & $1.30\pm0.16$\\
\shortgqlupb & 2006-09-18 & 00:02:15 & 00:07:15 & 25 & 300$\mathrm{\times1}$ & 0.84 & 1.40 & 1.96 & 13.12 & $1.30\pm0.16$\\
\shortgqlupb & 2006-09-18 & 00:07:41 & 00:12:41 & 25 & 300$\mathrm{\times1}$ & 0.90 & 1.42 & 1.92 & 13.01 & $1.30\pm0.16$\\
\shortgqlupb & 2006-09-18 & 00:24:36 & 00:29:36 & 25 & 300$\mathrm{\times1}$ & 0.73 & 1.52 & 2.23 & 13.67 & $1.30\pm0.16$\\
\shortgqlupb & 2006-09-18 & 00:30:08 & 00:35:08 & 25 & 300$\mathrm{\times1}$ & 0.75 & 1.55 & 2.18 & 13.76 & $1.30\pm0.16$\\
\shortgqlupb & 2006-09-18 & 00:47:01 & 00:52:01 & 25 & 300$\mathrm{\times1}$ & 0.95 & 1.67 & 1.76 & 14.44 & $1.30\pm0.16$\\
\shortgqlupb & 2006-09-18 & 00:52:33 & 00:57:33 & 25 & 300$\mathrm{\times1}$ & 1.17 & 1.72 & 1.43 & 14.81 & $1.30\pm0.16$\\
Hip078754 & 2006-09-18 & 01:34:42 & 01:36:42 & 25 & 120$\mathrm{\times1}$ & 1.08 & 1.95 & 1.59 & - & $1.30\pm0.16$\\
Hip087140 & 2006-09-18 & 01:17:36 & 01:19:36 & 25 & 120$\mathrm{\times1}$ & 1.17 & 1.22 & 1.30 & - & $1.30\pm0.16$\\
\hline

\shortgscb & 2017-04-25 & 06:21:07 & 06:26:07 & 100 & 300$\mathrm{\times1}$ & 0.54 & 1.01 & - & 16.44 & $1.46\pm0.30$\\
\shortgscb & 2017-04-25 & 06:32:12 & 06:37:12 & 100 & 300$\mathrm{\times1}$ & 0.52 & 1.00 & - & 16.15 & $1.46\pm0.30$\\
\shortgscb & 2017-04-25 & 06:48:51 & 06:53:51 & 100 & 300$\mathrm{\times1}$ & 0.60 & 1.00 & - & 16.35 & $1.46\pm0.30$\\
\shortgsca & 2017-04-25 & 07:05:42 & 07:05:57 & 100 & 15$\mathrm{\times1}$ & 0.46 & 1.00 & - & - & $1.46\pm0.30$\\
Hip079599 & 2017-04-25 & 07:17:56 & 07:17:57 & 100 & 1$\mathrm{\times1}$ & 0.52 & 1.01 & - & - & $1.46\pm0.30$\\
\hline
\shortgscb & 2017-04-25 & 07:35:50 & 07:40:50 & 100 & 300$\mathrm{\times1}$ & 0.60 & 1.02 & - & 16.30 & $1.46\pm0.30$\\
\shortgscb & 2017-04-25 & 07:46:53 & 07:51:53 & 100 & 300$\mathrm{\times1}$ & 0.63 & 1.03 & - & 16.46 & $1.46\pm0.30$\\
\shortgscb & 2017-04-25 & 08:03:31 & 08:08:31 & 100 & 300$\mathrm{\times1}$ & 0.66 & 1.05 & - & 16.50 & $1.46\pm0.30$\\
Hip095105 & 2017-04-25 & 08:32:45 & 08:32:48 & 100 & 3$\mathrm{\times1}$ & 0.57 & 1.05 & - & - & $1.46\pm0.30$\\
\hline
\shortgscb & 2017-05-04 & 04:29:49 & 04:34:49 & 100 & 300$\mathrm{\times1}$ & 0.48 & 1.10 & - & 16.18 & $1.62\pm0.33$\\
\shortgscb & 2017-05-04 & 04:40:54 & 04:45:54 & 100 & 300$\mathrm{\times1}$ & 0.56 & 1.07 & - & 16.06 & $1.62\pm0.33$\\
\shortgscb & 2017-05-04 & 04:57:38 & 05:02:38 & 100 & 300$\mathrm{\times1}$ & 0.46 & 1.05 & - & 16.00 & $1.62\pm0.33$\\
\shortgsca & 2017-05-04 & 05:14:44 & 05:14:59 & 100 & 15$\mathrm{\times1}$ & 0.62 & 1.03 & - & - & $1.62\pm0.33$\\
Hip081214 & 2017-05-04 & 05:27:03 & 05:27:05 & 100 & 2$\mathrm{\times1}$ & 0.86 & 1.08 & - & - & $1.62\pm0.33$\\
\hline
\shortgscb & 2017-06-01 & 01:54:27 & 01:59:27 & 100 & 300$\mathrm{\times1}$ & 0.86 & 1.21 & 7.73 & 16.09 & $1.28\pm0.33$\\
\shortgscb & 2017-06-01 & 02:16:29 & 02:21:29 & 100 & 300$\mathrm{\times1}$ & 0.74 & 1.15 & 11.25 & 15.88 & $1.28\pm0.33$\\
\shortgscb & 2017-06-01 & 02:27:32 & 02:32:32 & 100 & 300$\mathrm{\times1}$ & 0.66 & 1.12 & 10.24 & 15.98 & $1.28\pm0.33$\\
\shortgscb & 2017-06-01 & 02:44:12 & 02:49:12 & 100 & 300$\mathrm{\times1}$ & 0.72 & 1.09 & 10.56 & 15.91 & $1.28\pm0.33$\\
\shortgsca & 2017-06-01 & 03:01:00 & 03:01:15 & 100 & 15$\mathrm{\times1}$ & 0.85 & 1.06 & 14.31 & - & $1.28\pm0.33$\\
Hip084435 & 2017-06-01 & 03:16:58 & 03:17:04 & 100 & 6$\mathrm{\times1}$ & 0.60 & 1.13 & 13.61 & - & $1.28\pm0.33$\\
Hip084435 & 2017-06-01 & 06:11:06 & 06:11:14 & 100 & 8$\mathrm{\times1}$ & 0.65 & 1.03 & 16.46 & - & $1.28\pm0.33$\\
\hline
\shortgscb & 2017-08-19 & 01:28:50 & 01:33:50 & 100 & 300$\mathrm{\times1}$ & 0.64 & 1.18 & 4.58 & 16.34 & $1.80\pm0.35$\\
\shortgscb & 2017-08-19 & 01:39:28 & 01:44:28 & 100 & 300$\mathrm{\times1}$ & 0.66 & 1.21 & 4.67 & 16.39 & $1.80\pm0.35$\\
\shortgscb & 2017-08-19 & 01:55:28 & 02:00:28 & 100 & 300$\mathrm{\times1}$ & 0.70 & 1.27 & 4.90 & 16.84 & $1.80\pm0.35$\\
\shortgsca & 2017-08-19 & 02:11:40 & 02:11:55 & 100 & 15$\mathrm{\times1}$ & 0.71 & 1.32 & 6.09 & - & $1.80\pm0.35$\\
Hip082670 & 2017-08-19 & 01:20:24 & 01:20:26 & 100 & 2$\mathrm{\times1}$ & 0.57 & 1.18 & 5.48 & - & $1.80\pm0.35$\\
\hline

\label{tab:obs}
\end{longtable}
}

\subsection{Monitoring sequences with VLT/SINFONI}
We monitored \shortgqlupb\ and \shortgscb\ with the Integral Field Spectrograph (IFS) SINFONI.  SINFONI is a now decommissioned instrument previously installed at VLT/UT4 Cassegrain focus and composed of a
custom adaptive optics module (MACAO) feeding an IFS (SPIFFI). SPIFFI splits the field-of-view into 32 horizontal slices (slitlets) and re-arranges them to form a pseudo-slit which is dispersed by a grating on a 2k$\times$2k HAWAII-2RG detector \citep{Eisenhauer2003SINFONIVLT, Bonnet2004AstronomicalInstrumentation}.  The instrument was operated with pre-optics and gratings in the J-band (1.1 -- 1.35~\um). 
For \shortgqlupb, the setup provided a 0.8" square field of view (FoV) composed of $12.5 \times 25$~mas spaxels, with a resolving power of $R_{\uplambda}=\frac{\uplambda}{\Delta\uplambda}=2360$.
For \shortgscb, the setup provided a 3.2" square field of view (FoV) composed of $50 \times 100$~mas spaxels, with a resolving power of $R_{\uplambda}=\frac{\uplambda}{\Delta\uplambda}=1800$.
They translate into Line Spread Function (LSF) widths of 127 and 167~\kms\ for \shortgqlupb\ and \shortgscb\ respectively.

MACAO used the host stars for wavefront sensing. The de-rotator at the Cassegrain focus was used to align the star and companions horizontally in the FoV at acquisition. A telescope offset was then applied to place the host stars PSF cores outside of the FoV and allow for deeper exposures. Small additional dithering were applied along the vertical direction of the FoV in-between each exposure to ensure that the companion signal did not fall all the time on bad pixels. A blank-sky exposure (with identical DIT) was performed at the end of the dithering sequence to ensure a proper removal of the OH- emission lines and thermal background. We then moved the star back in the FoV and recorded an additional short-exposure to record a PSF, to be used for calibration purposes.

\shortgscb\ was observed on April 25, May 4, June 1, and August 18, 2017 as part of our program (099.C-0760). \shortgqlupb\ was observed on April 25, 30, 2017 and June 1, 2017 as part of the same program (UT dates). We also decided to re-process SINFONI data of \shortgqlupb\ obtained on 2006-09-18 (program 077.C-0264) and presented in \cite{2007A&A...463..309S} for consistency and given that the published SINFONI spectrum has been shown to be discrepant with respect to other studies \citep{mcelwain_first_2007, lavigne_near-infrared_2009}. These observations used the same setup and followed an identical data-extraction strategy as for the 2017 observations. The log of observations is detailed in Table~\ref{tab:obs}.

\subsection{SINFONI cube reconstruction and calibration} \label{sec:cube_reconstruction}

SINFONI records bi-dimensional raw frames  of the 32 dispersed slitlets. We used the \textit{Toolkit for Exoplanet deTection and chaRacterization with IfS} (hereafter \texttt{TExTRIS}) \citep[][Bonnefoy et al., in prep]{2021A&A...648A..59P} to correct them from static noise caused by the detector electronics \citep{Bonnefoy14}. 

The SINFONI data handling pipeline v3.0.0 \citep{2006NewAR..50..398A} was then used to reconstruct calibrated data cubes encoding the spatial and spectral dimensions (X, Y, $\uplambda$) from these raw data. The pipeline uses calibration frames obtained at day time to apply basic cosmetic steps on the raw frames and correct them from the distortion introduced by optical elements and the detector. It identifies the position of slitlets on the raw frames and builds a reference 2D map associating the pixels to given wavelengths. The sky emission was evaluated and removed through the field subtracting cubes of reconstructed exposures on sky. Sky OH- emission lines residuals were then evaluated and subtracted using the method of \cite{2007MNRAS.375.1099D}. Because residuals hot/cosmics pixels were found in the cubes, they were corrected for using the \textsc{lacosmic} Python package on each individual wavelength frame \citep{vandokkum_cosmicray_2001}.

The wavelength solution relies on day-time calibration frames obtained with an Argon lamp and has proven to be inaccurate on other datasets \citep{2021A&A...648A..59P, 2021Natur.595..370Z}. We therefore improved the pipeline wavelength calibration with routines from \texttt{TExTRIS}, comparing the many telluric absorption lines contained in each spaxel to a model generated for the observing conditions using the ESO Skycalc tool \citep{refId0, refId1}.
The method is applicable whenever the spaxels contain enough stellar light to infer the telluric absorptions. We evaluated the shifts in the wavelength interval 1.1--1.3~\um\ using cross-correlations of spaxels with a signal-to-noise ratio (S/N) above 10 at \PaBeta. A master median-combined map of wavelength shift was then created and each individual cube was corrected from the median of the shifts of individual spaxels meeting the S/N criterion. The corrected wavelength shifts at each epochs are reported in Table~\ref{tab:obs}, for an average wavelength calibration uncertainty of $\sim0.2$~\AA\ and $\sim0.3$~\AA\ for \shortgqlupb\ and \shortgscb\ respectively. Because of the relatively low S/N in the science target cubes, we used the  wavelength solution found with the telluric standard stars (relying on the same original calibration as the science target) as the solution for the whole sequence.

\subsection{New investigation of Keck/OSIRIS data}
We also include in the analysis the individual spectra of \shortgscb\ obtained on July 9, 2010 with the OH-Suppressing Infrared Imaging Spectrograph \citep[OSIRIS;][]{2006SPIE.6269E..1AL} mounted on Keck II. OSIRIS is a lenslet-based IFS operated with the Keck adaptive optics system. The instrument produced data cubes sampling a rectangular FoV ($0.8\times3.2$") with 0.05x0.05$^{\prime\prime}$ spaxels. The J$_{\textrm{bb}}$ band grating provided 1.18-1.416~\um\ spectra at R$\sim$3800.  Eight 5-minutes long exposures were recorded. The stacked data cubes are presented in \cite{bowler_disk_2011} and the companion is sufficiently separated from the star to directly extract its spectrum within a circular aperture. The data analysis is described in more details in \cite{bowler_disk_2011}. We re-analyzed here the individual data cubes in an attempt to detect \PaBeta\ variability from one exposure to the next.


\section{Spectral extraction}
\label{sec:extraction}

\subsection{Halo removal}


Both \shortgqlupb\ and \shortgscb\ spectra are contaminated by the PSF wings of the  primary stars located outside of the FoV. The data cubes obtained at the end of each sequence of observations with the host stars in the FoV do not probe the extended wings of the PSF and therefore do not allow for direct subtraction of the contamination. We therefore had to develop a specific procedure described below and applied to each individual SINFONI data cube on both targets.

\begin{figure}
\centering
\includegraphics[width=\columnwidth]{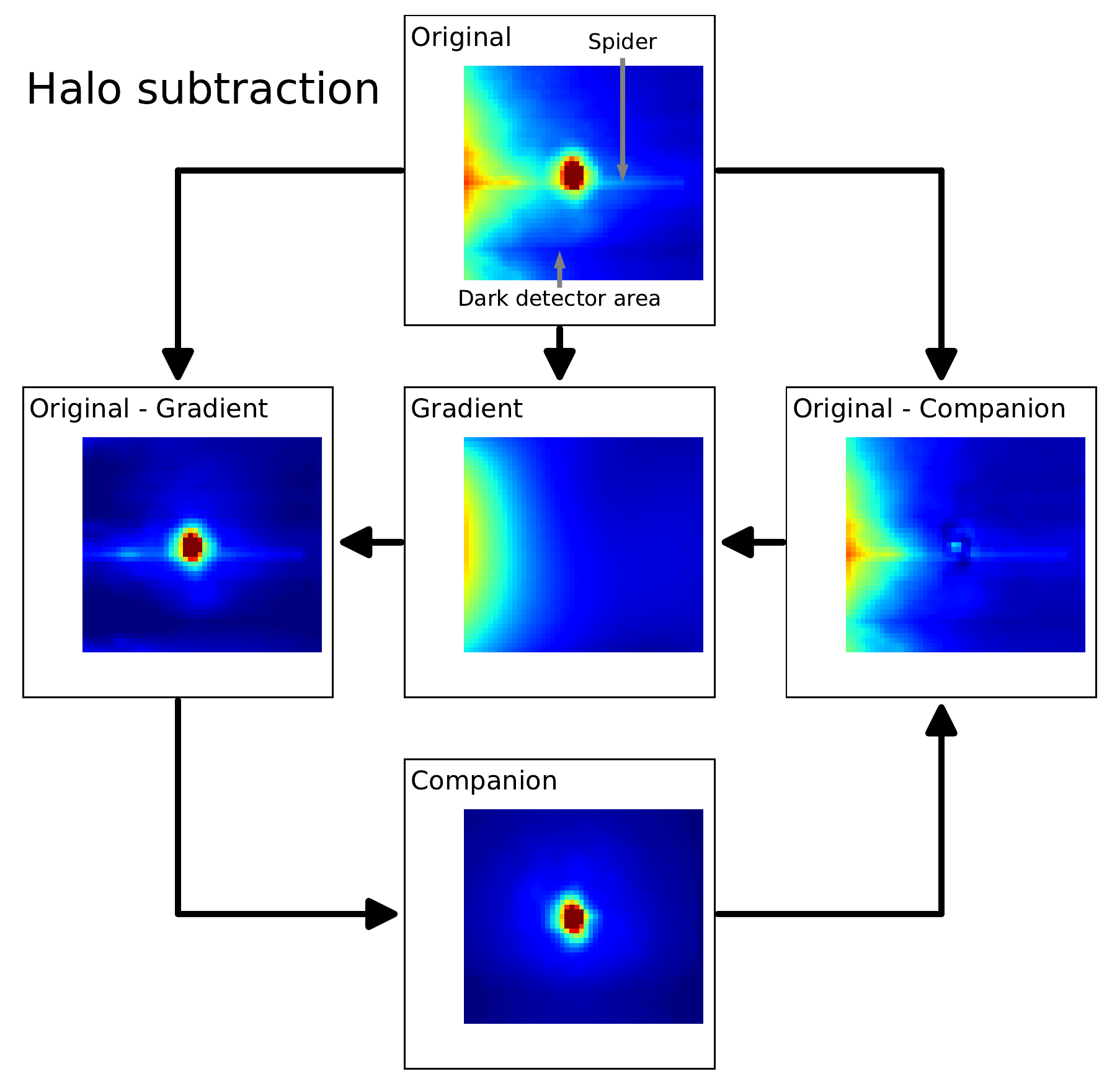}

\caption{Illustration of the halo subtraction, for the first data cube of the \gqlupb\ 2017-04-30 epoch (median along the wavelength axis). We also report the location of the diffraction pattern of the telescope spiders, along with the identified area with improperly calibrated instrument transmission that affects the lower-half part of the images.}
\label{fig:clean_performance}
\end{figure}

Before halo removal, the cube rectangular spaxels were linearly interpolated to produce a regular spatial sampling of 12.5~mas in the horizontal and vertical directions. The contamination was then evaluated and corrected in each individual frame of the data cubes, with a CLEAN-like algorithm \citep{1974A&AS...15..417H}. The algorithm fits a two component flux model made of: (i) a third-order bi-dimensional gradient, approximating the primary star PSF wings, (ii) a PSF model placed at the companion location and scaled to the companion's flux. The latter step requires the definition of a PSF model. We used the primary star data cubes (normalized to 1) for that purpose in the 2017 data. Data cubes of the primary star were not available in the 2006 data and we chose to use those corresponding to the telluric standard star instead. The two flux components were successively fitted and subtracted from the original frames, for a total of 4 iterations, which is a trade-off between $\chi^2_r$ and S/N optimization. See Fig.~\ref{fig:clean_performance} for a visual representation, and Fig.~\ref{fig:appendix:residuals} for residuals.

The host star gradient subtraction allows us to subtract the large-scale structures but can not remove more complex features such as the diffraction pattern caused by the telescope spiders. This component is most important on \shortgqlupb\ and was then approximated as a 1-d Gaussian profile in the direction perpendicular to the spider axis and subtracted in the gradient-subtracted cubes.

SINFONI shows a static dark horizontal area in the lower half part of all data cubes (see Fig~\ref{fig:clean_performance}) likely caused by persistence appearing in the lamp flat-fields taken at day time. This was discovered during data reduction and affected the spectra profiles when the companion was located in this region. We decided to exclude the corresponding cubes and spectra from the analysis.

\begin{figure*}[t]
    \centering
    \includegraphics[width=\textwidth]{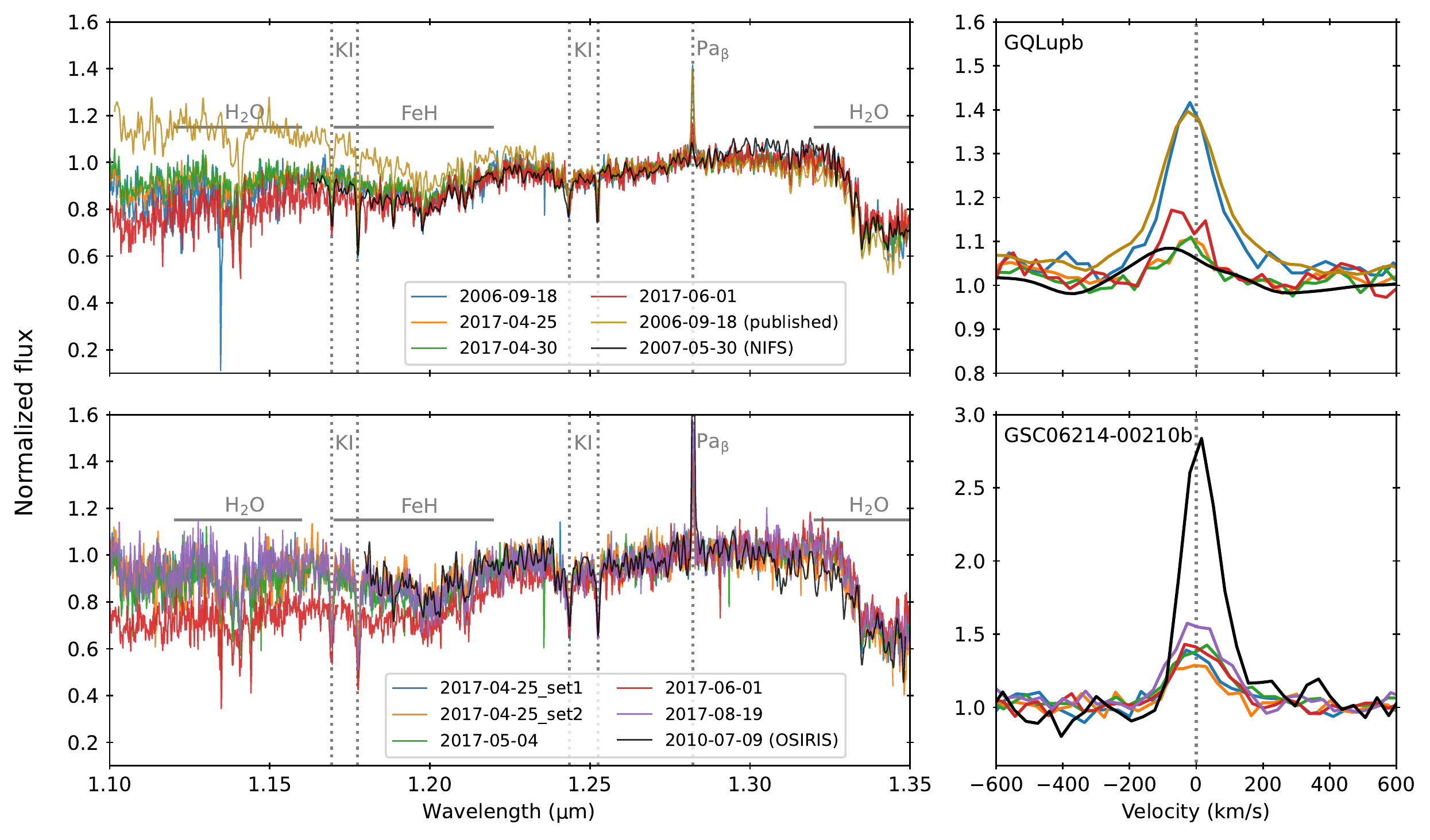}
    
    \caption{Companion spectra. Top row: \gqlupb\ spectra. Bottom row: \gsc\ spectra. Left column: full spectra. Right column: zoom on the \PaBeta\ line. In both cases, the spectra continuum are normalized to 1 as the mean value on the 1.27--1.28~\um\ and 1.29--1.30~\um\ ranges. The 2010-07-09 (OSIRIS) epoch of \shortgscb\ was degraded to SINFONI resolution (R=1800) in the left column, but is shown at full resolution (R=3800) in the right column. The spectral slopes differences are the results of different Strehl Ratios between epochs, despite the AO correction. Our re-extraction of \gqlupb\ spectrum initially presented in \cite{2007A&A...463..309S} produces a pseudo-continuum short-wards of 1.24~\um\ in better agreement with the more recent SINFONI epochs and observations produced by NIFS.}
    \label{fig:spectra}
\end{figure*}
\begin{figure*}
    \centering
    \includegraphics[width=\textwidth]{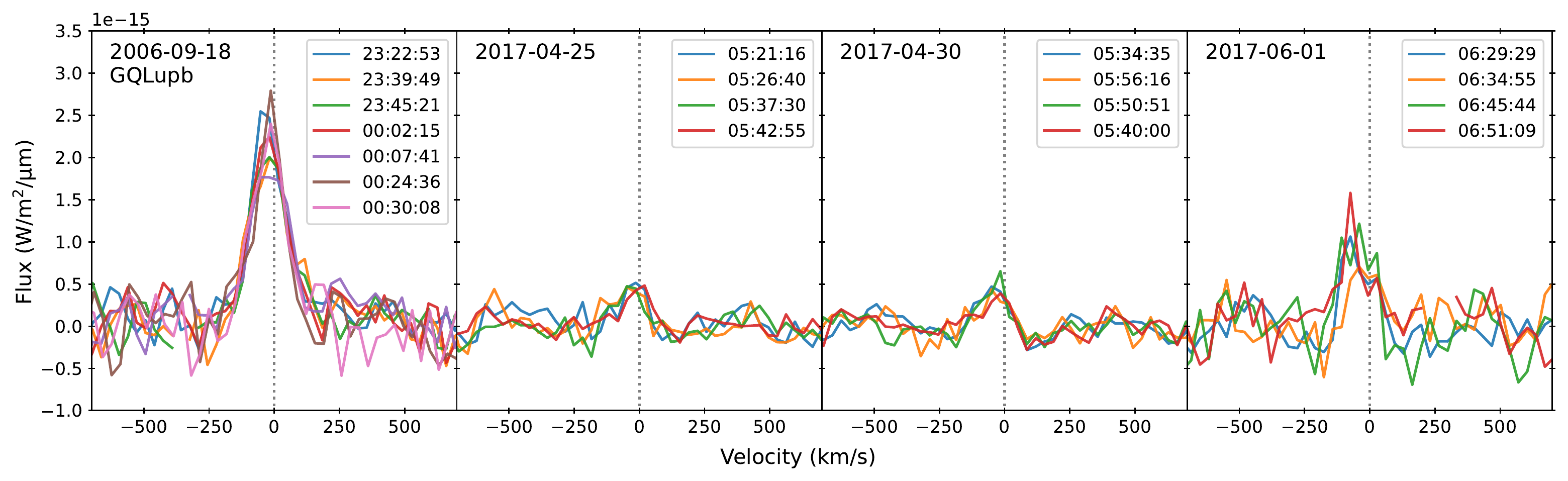}\\
    \includegraphics[width=\textwidth]{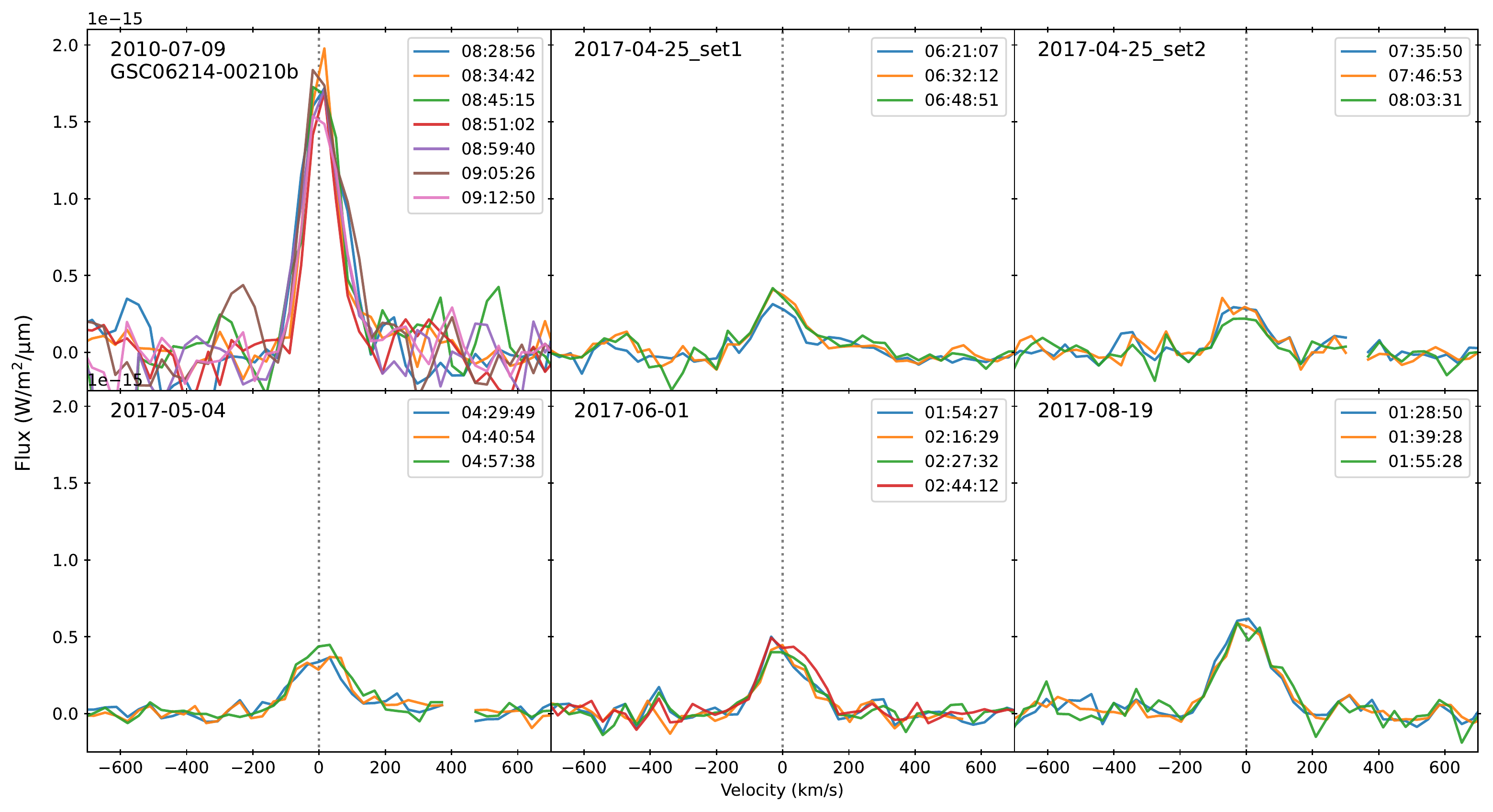}
    \caption{\PaBeta\ profiles, corrected from the atmospheric contribution and baseline subtracted (see Sects.~\ref{sec:results:photosphere_fitting} \& \ref{sec:results:analysis_pabeta}). Each spectrum is labeled with the corresponding UT time at the start of the integration. First row: \shortgqlupb\ lines for each epoch. Second and third rows: same for \shortgscb, with the addition of the individual OSIRIS lines from \cite{bowler_disk_2011} (2010-07-09 epoch).}
    \label{fig:individual_lines}
\end{figure*}

\subsection{Extraction and flux calibration}
\label{subsec:fluxcal}

The PSF model estimated from the primary (or telluric standard) star was not a good enough match to the companion PSF to extract spectra from profile fitting. Therefore, the companions spectra were extracted within a circular aperture on the contamination-corrected data cubes of constant radius with wavelengths.
The size of the aperture  was chosen to conserve the spectral slope and to avoid integrating noise due to the imperfect removal of the stellar halo in the background-limited area. This procedure is similar as the one adopted in \cite{Bonnefoy14} and proven to provide a robust extraction of the spectral features and slope on J-band SINFONI data. The adopted radii were ranging from 50 to 75mas. We retrieve spectra of various S/N depending on the observing conditions. For \shortgqlupb, the S/N ranges from 20 to 30 in the 2006-09-18 and 2017-06-01 epochs, while it ranges from 35 to 50 in the 2017-04-25 and 2017-04-30 epochs. For \shortgscb, the S/N is relatively stable, within the 15-30 range for all epochs.

The atmospheric transmission was evaluated computing the ratio between the extracted spectra of telluric standard stars and their synthetic spectra: black-bodies scaled to the 2MASS JHK, and WISE W1 and W2 photometry of the objects \citep{2003yCat.2246....0C, 2012yCat.2311....0C} collected with the VOSA\footnote{\url{http://svo2.cab.inta-csic.es/theory/vosa/}} tool \citep{2008A&A...492..277B}. The \PaBeta\ absorption line of the standards were evaluated and removed beforehand as follows. The telluric standard star spectrum was corrected from telluric absorptions by a model created with the SkyCalc tool \citep{refId0, refId1}. The line profile was then fitted in the by a Voigt profile function, then subtracted from the spectrum. The spectrum was then applied the same telluric absorption model. This ensures that the line shape evaluation is not impacted by weak telluric features in the line wings and that it does not bias the companion emissions at these wavelengths.

We performed an absolute flux calibration of the individual spectra using the telluric standard star as a spectro-photometric reference (2MASS J-band photometry). The use of the telluric standard stars was favored over the use of the host stars because of their potential photometric variability. This is the case for \gqlupa\ which is a known high-amplitude variable star \citep[$\Delta J=0.87$,][]{2001ApJ...548..377B, 2007A&A...468.1039B}. The remaining J-band magnitudes inferred from the individual epochs were averaged to remove part of the uncertainty on the photometric calibration due to  variation in observing conditions and AO performance between the telluric standard stars and the companion observations. We decided to exclude the 2006 epoch on \gqlupb\ from this process because the companion and the telluric standard were observed under poor and variable adaptive-optics conditions.

We find a magnitude J = $16.21^{+0.26}_{-0.21}$ on \shortgscb. The value is well consistent with the one measured by \cite{lachapelle_characterization_2015} and \cite{bowler_disk_2011} (J=$16.19 \pm 0.04$ and J$=16.25 \pm 0.04$, respectively) derived from the contrast between the primary and the companion, and the tabulated 2MASS J-band magnitude of the system. We decided to use the latter value.

We find J~=~$14.38^{+0.13}_{-0.12}$ for \gqlupb. \cite{lavigne_near-infrared_2009} found J~=~15.13 $\pm$ 0.44 but the measurement accuracy was limited by the partial saturation of the host star in their data and the photometry relies on the 2MASS magnitude of the host star which is known to be variable, as previously mentioned. \cite{mcelwain_first_2007} found J~=~14.90 $\pm$ 0.11 using a similar procedure as ours and this photometry was used in all contemporaneous studies on the system \citep[e.g.,][]{2021AJ....162..286S}. But this measurement was performed on OSIRIS data at high airmass with a narrower wavelength range and at a lower signal-to-noise ratio. Given the amplitude of variability of the primary star \citep[$\Delta$J~=~0.87,][]{2007A&A...468.1039B}, the discrepancy between published values and ours is not surprising.
The SINFONI photometry makes the J-K color of \shortgqlupb\ \citep[$1.0 \pm 0.2$~mag, assuming the K-band magnitude of \gqlupb re-evaluated by ][]{2008A&A...484..281N} in better agreement with that of other companions of the same luminosity, spectral type, and age range \citep[e.g., Fig.~4 of][]{2018ApJ...854L..27G}. Furthermore, our value is also consistent with the magnitude correction (0.30 to 0.66~mag) found by \cite{2021AJ....162..286S} to adjust with a single atmospheric model the J-band scaled to the \cite{mcelwain_first_2007} photometry with the rest of the companion spectral energy distribution. We therefore choose to adopt the J-band magnitude inferred from the SINFONI data later on (see Sect.~\ref{sec:discussion:continuum_variability} for a discussion on possible photometric variations).

Spectra are also corrected for the Barycentric Earth Radial Velocity (BERV), the primary radial velocities, and the companion keplerian motions. The BERV is computed following \cite{piskunov_new_2002} for each individual data cube. \shortgqlupb\ radial velocity is $-0.8 \pm 0.4$~\kms\ \citep{2016A&A...593A..74S}. \shortgscb\ radial velocity is $-3.98 \pm 0.88$~\kms, it was computed from a combination of its orbital velocity ($0.61 \pm 0.09$~\kms, \cite{2019AJ....157...71P}) and the radial velocity of the primary star ($-4.59 \pm 0.88$~\kms, GAIA DR2). Radial velocities are given in the heliocentric system.

For all the following analysis, error bars were computed from the dispersion  around \PaBeta, on the residual photosphere-subtracted and baseline-corrected (polynomial approximation of residual baselines, see Sect.~\ref{sec:results:analysis_pabeta}) spectra. For velocity conversion (e.g., right panels of Fig.~\ref{fig:spectra}), we used the \PaBeta\ rest-frame wavelength gathered on NIST \citep{NIST_ASD}, $\uplambda_0 = 1.2821578$~\um \citep[vacuum,][]{1996SoPh..165..257C}.

\subsection{Photospheric contribution}
\label{sec:methods:photosphere_fitting}
Both \shortgqlupb\ and \shortgscb\ exhibit a rich spectrum of molecular absorption lines and shallow KI doublets at 1.169/1.177~\um\ and 1.243/1.253~\um\ characteristic of young  M-L transition objects \citep[e.g.,][]{Bonnefoy14}. Molecular absorption lines overlap with the \PaBeta\ line, affecting the emission line profiles. The spectra were therefore corrected for the photospheric features using our forward modeling tool \formosa\ \citep{petrus_new_2020, 2023A&A...670L...9P}.

\formosa\ compares pre-computed grids of atmospheric models with observations using the Nested Sampling Bayesian inference algorithm \citep{2004AIPC..735..395S}. We used as input the ATMO models \citep{tremblin_fingering_2015,tremblin_cloudless_2016} which explore effective temperature and surface gravity in the expected range of values for the late-M companions. The grids are described in more details in Table C.1 of \cite{2023A&A...670L...9P} and allow to explore non-solar C/O and [M/H] values, as well as the adiabatic index $\upgamma$. \formosa\ also allows for fitting of the radial velocity (RV), rotational broadening (vsin(i)),  limb-darkening coefficient ($\upepsilon$), and interstellar extinction (Av). We did not fit the photospheric contribution on each individual spectrum, but rather on the averaged spectrum at each epoch and used the RV shift to compensate for residual wavelength shifts from one epoch to another that may arise from imperfect wavelengths re-calibration.

\begin{figure*}[ht]
    \centering
    \includegraphics[width=0.48\linewidth]{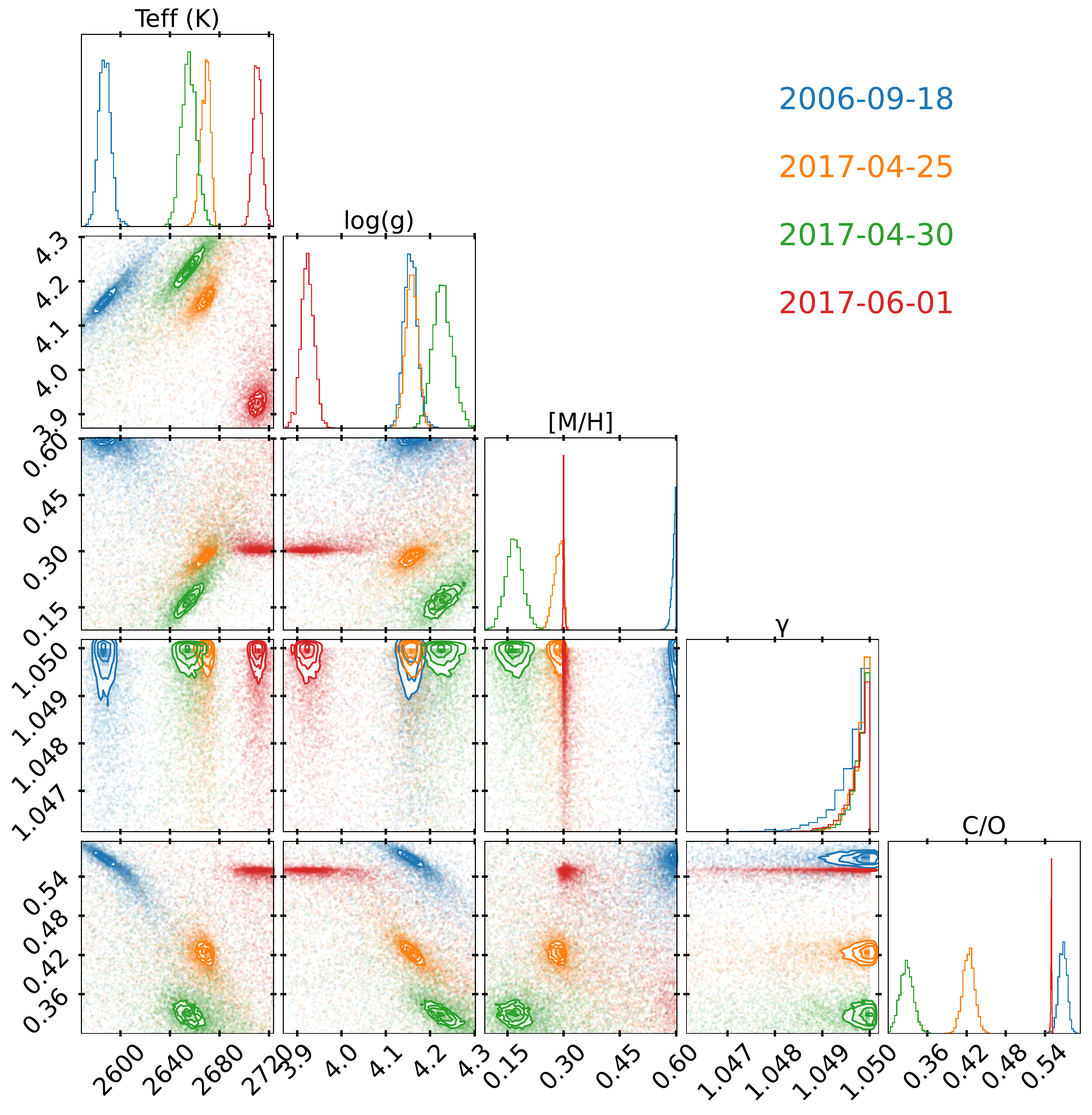}
    \includegraphics[width=0.48\linewidth]{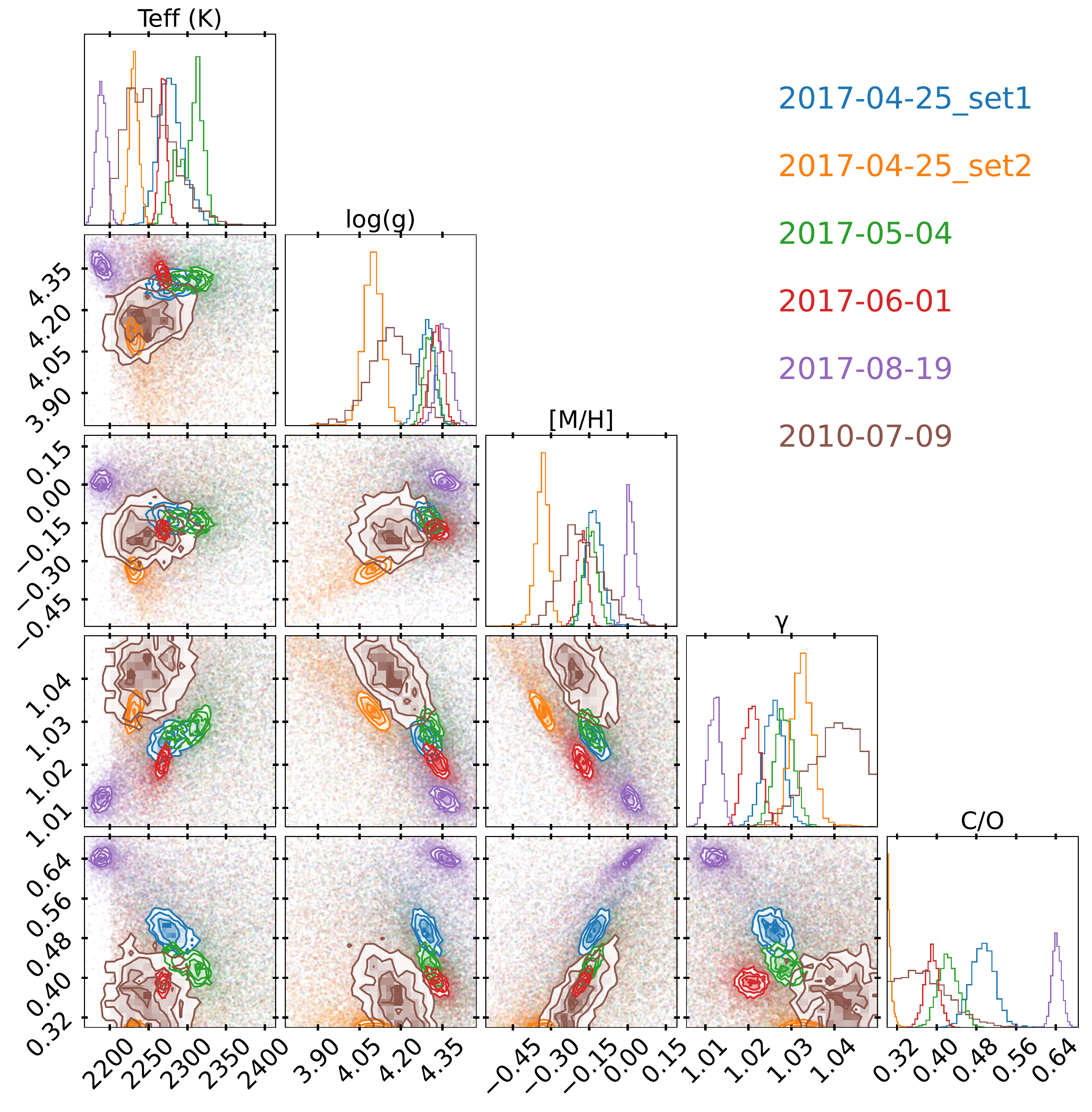}
    \caption{\formosa\ resulting photosphere parameters. Left: \shortgqlupb\ results. Right: \shortgscb\ results. The Radius, Av, radial velocity (RV) and limb-darkening coefficients are omitted from these plots for the sake of clarity.}
    \label{fig:results:formosa_corner_plots}
\end{figure*}

The resulting parameter corner plots are presented in Fig.~\ref{fig:results:formosa_corner_plots}. The 1D-histograms were used to find the resulting best values for each parameter, bi-modal Gaussian fitting (different $\upsigma$ for the upper and lower halves). The free parameters are the peak location $\upmu_0$ and the two sigmas $\upsigma_1$ and $\upsigma_2$. The retained value is $\upmu_0$ with the corresponding error bars. 
The dispersion of values between epochs makes it clear that the Bayesian process on individual epochs provides posterior distributions whose extents are not representative of the true error bars on the values. We therefore estimate the adopted value as the mean of all epochs, and error bars from the dispersion of the ensemble of posteriors of the different epochs. Values at the edge of the grid (upper/lower limits) were ignored when computing the adopted value.
The results are presented in Table~\ref{tab:formosa_results} and discussed in the following section.

\begin{table*}[ht]
\caption{Atmospheric parameters retrieved with \formosa}             
\label{tab:formosa_results}      
\begin{center}
\footnotesize
\begin{tabular}{c l c c c c c c c c}
\hline\hline       
& Date & Teff & log(g) & [M/H] & $\upgamma$ & C/O & Radius & $A_v$ & RV\\ 
& YYYY-MM-DD & (K) &  &  &  &  & (\Rjup) &  & (km/s)\\ 
\hline
\multirow{5}*{\rotatebox{0}{\shortgqlupb}}
& 2006-09-18&$2587$ &$4.16$ &$\footnotemark[1]>0.59$ &$\footnotemark[1]>1.05$ &$0.57$ &$4.45$ &$2.38$ &$3.03$ \\ 
& 2017-04-25&$2671$ &$4.16$ &$0.30$ &$\footnotemark[1]>1.05$ &$0.42$ &$4.08$ &$2.13$ &$5.59$ \\ 
& 2017-04-30&$2655$ &$4.22$ &$0.17$ &$\footnotemark[1]>1.05$ &$0.33$ &$4.07$ &$1.91$ &$13.59$ \\ 
& \footnotemark[2]2017-06-01&$2711$ &$3.92$ &$0.30$ &$\footnotemark[1]>1.05$ &$0.55$ &$\footnotemark[1]>5.00$ &$4.32$ &$13.11$ \\ 
& adopted & $2638^{+33}_{-51}$ & $4.18^{+0.04}_{-0.02}$ & $0.23^{+0.06}_{-0.06}$ & >1.05 & $0.44^{+0.13}_{-0.11}$ & $4.20^{+0.25}_{-0.13}$ & $2.14^{+0.24}_{-0.23}$ & - \\

\hline

\multirow{7}*{\rotatebox{0}{\shortgscb}}
& 2017-04-25\_set1&$2274$ &$4.30$ &$-0.13$ &$1.03$ &$0.50$ &$1.45$ &$0.45$ &$-18.06$ \\ 
& 2017-04-25\_set2&$2230$ &$4.10$ &$-0.33$ &$1.03$ &$\footnotemark[1]<0.30$ &$1.47$ &$\footnotemark[1]<0.00$ &$-8.78$ \\ 
& 2017-05-04&$2317$ &$4.30$ &$-0.15$ &$1.03$ &$0.42$ &$1.56$ &$1.50$ &$-12.73$ \\ 
& \footnotemark[2]2017-06-01&$2269$ &$4.33$ &$-0.18$ &$1.02$ &$0.39$ &$2.45$ &$\footnotemark[1]>4.99$ &$-19.58$ \\ 
& 2017-08-19&$2189$ &$4.36$ &$0.00$ &$1.01$ &$0.64$ &$1.58$ &$\footnotemark[1]<0.00$ &$-6.51$ \\ 
& 2010-07-09&$2232$ &$4.17$ &$-0.21$ &$1.04$ &$0.36$ &$1.37$ &$\footnotemark[1]<0.00$ &$-11.90$ \\ 
& adopted & $2248^{+69}_{-59}$ & $4.24^{+0.11}_{-0.15}$ & $-0.16^{+0.17}_{-0.17}$ & $1.03^{+0.01}_{-0.02}$ & $0.48^{+0.16}_{-0.12}$ & $1.49^{+0.10}_{-0.12}$ & $0.97^{+0.53}_{-0.53}$ & - \\ 

\hline

\end{tabular}
\end{center}

\tablefoot{Adopted values are obtained as described in Sect.~\ref{sec:methods:photosphere_fitting}. \footnotetext[1]{Unconstrained parameter, fit converged at the edge of the parameter range.}\footnotetext[2]{Epoch ignored when computing the adopted value} }

\end{table*}

\begin{figure*}
    \includegraphics[width=\textwidth]{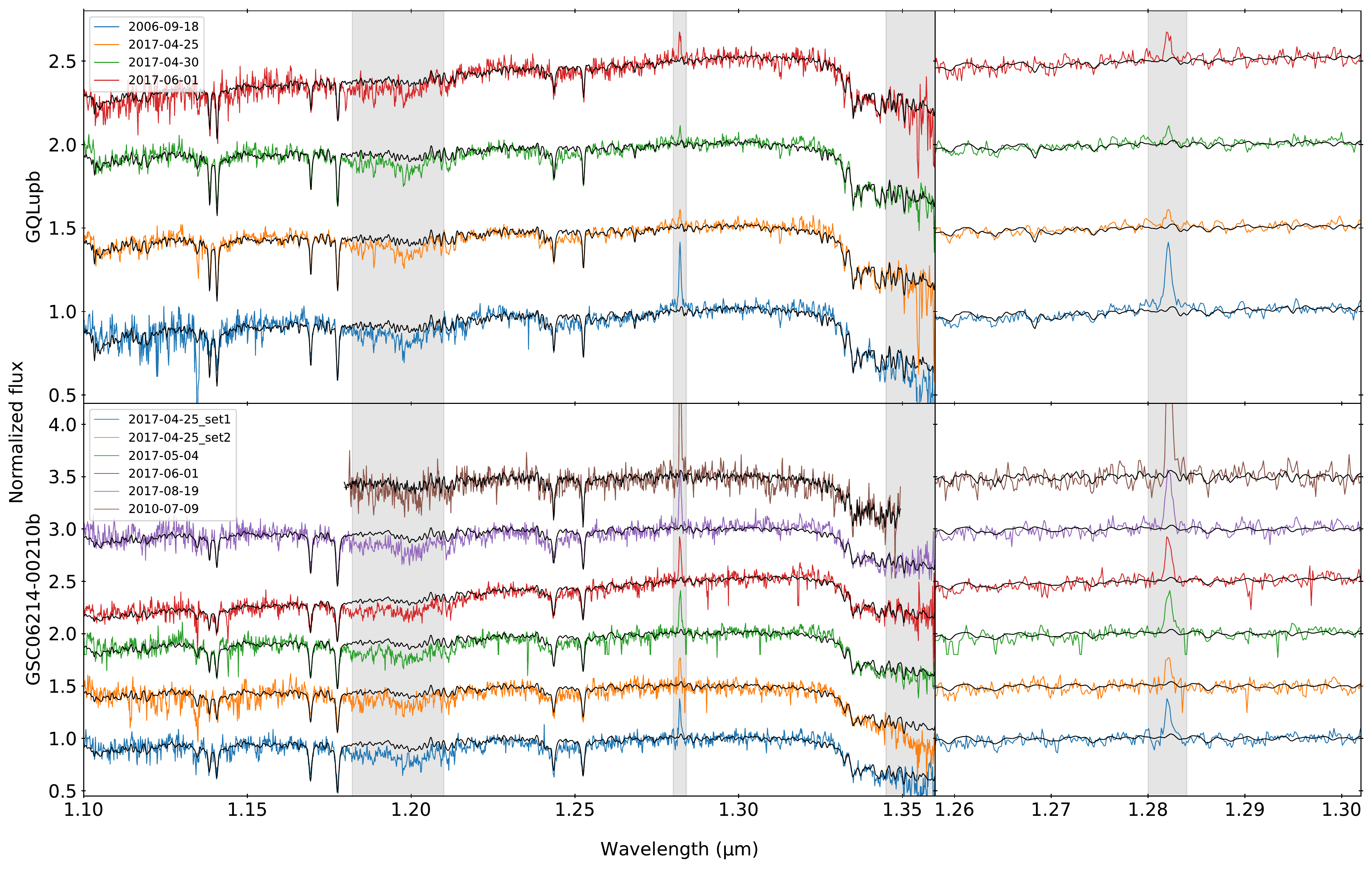}
    \caption{Visual comparison of the photosphere fitting with \formosa, for each epoch of both objects, with a zoom on \PaBeta\ in the right column. Grayed out wavelength ranges were excluded from fitting as the model either did not include corresponding species (FeH absorption band, \PaBeta) or data were too noisy ($>$1.34~\um).}
    \label{fig:appendix:formosa_visual}
\end{figure*}


\section{Results} \label{sec:results}

In this section, we present the resulting spectra and atmospheric parameters, along with their lines properties (profiles and variability) and modeling.

\subsection{Companion J-band spectra}

The mean spectra for each epoch are presented in Fig.~\ref{fig:spectra}, along with a zoom on the \PaBeta\ lines. \PaBeta\ lines of individual data cubes are presented in Fig.~\ref{fig:individual_lines}. The spectra all have a relatively similar spectral slope, with a discrepancy for the 2017-06-01 epoch for both objects. We argue that this is due to the lower Strehl ratios during that night -- the AO performance was significantly worse than for other epochs -- which is affecting mostly the blue part of the spectrum, leading to larger flux loss at smaller wavelengths.
Overall, our spectral slope is in agreement with the already published spectra \citep{lavigne_near-infrared_2009, bowler_disk_2011}, but the \shortgqlupb\ spectra differ from the \cite{2007A&A...463..309S}, which was already pointed out by \cite{lavigne_near-infrared_2009}. This may be explained by residual primary flux in their spectrum, which is in line with the wider \PaBeta\ line profile (See Fig.~\ref{fig:spectra}, top-right panel). 
We observe spectroscopic features typical of late-M/early-L dwarfs for both objects at all epochs: the two KI doublets (1.17 and 1.25~\um), the FeH absorption band (1.17--1.22~\um) and the numerous H$_{2}$O absorption lines (1.13--1.16~\um\ and $>$1.32~\um, see Fig.~\ref{fig:spectra}). The spectra are also composed of numerous less prominent molecular features all over the band, which are best described by photosphere models (see Section~\ref{sec:results:photosphere_fitting}).

\subsection{Atmospheric and physical parameters} \label{sec:results:photosphere_fitting}

We present here the results from atmospheric modeling of the J-Band spectra with \formosa\ (Sect.~\ref{sec:methods:photosphere_fitting}).
Best fits for both sources can be found in Table~\ref{tab:formosa_results}. A visual comparison of data and fitted spectra can be found in Fig.~\ref{fig:appendix:formosa_visual}. A comparison of the various atmospheric parameters can be found in Fig.~\ref{fig:results:formosa_corner_plots}, in the form of corner plots.
Some parameters converged at the edge of the parameter range, such as the adiabatic index $\upgamma$, either consistently or occasionally. The spectral slopes of both companions on June 1, 2017 are discrepant with the other epochs and the retrieved atmospheric parameters (at the very least \Teff and Av) are likely non representative.

\subsubsection{Inferred values and correlations}

Correlations are observed between \Teff and log~g. The behavior is similar to that observed on the spectral analysis of the young late-M companion  HIP~78530B with \formosa\ in \cite{petrus_new_2020} (albeit they used BT-SETTL atmospheric models). The correlation might be caused by the increase in strength of the K I and FeH absorptions with both decreasing \Teff and increasing surface gravities at the M-L transition \citep{2010ApJS..186...63R, Bonnefoy14}. 

On the other hand, the log~g and C/O are anti-correlated. This is due to the sensitivity to these two parameters of the numerous FeH and H$_{2}$O absorptions at J-band. Finally, both sources show either no correlations between [M/H] and C/O, or a positive one. We do not notice strong degeneracies between $\upgamma$ and the others parameters, which suggests that the limited exploration of that parameter does not bias the estimate of the other parameters.

For \shortgqlupb, atmospheric parameters are consistent for all epochs: $\textrm{log g} \sim 4.2$, $\textrm{T}_{\textrm{eff}}=2638^{+33}_{-51}$~K (beware that error bars do not include systematics of the model) and $\upgamma>1.05$. The log~g and \Teff are both well consistent with the values found in \cite{2021AJ....162..286S} adjusting the VLT/MUSE optical spectrum of the companion with BT-SETTL models. We however find with \formosa\ a higher radius than the estimates of \cite{2021AJ....162..286S} (found by adjusting a synthetic spectra on the companion's spectro-photometry). This might be due to the increased J-band apparent flux that we measure in Section \ref{subsec:fluxcal}. Part of the difference could also be caused by the use of different atmospheric models (see below).

For \shortgscb, we find $\textrm{log g}\sim 4.24$, $\textrm{T}_{\textrm{eff}}=2248^{+69}_{-59}$ K and $\upgamma=1.03^{+0.01}_{-0.02}$. \cite{lachapelle_characterization_2015} found $\textrm{T}_{\textrm{eff}}=2300 \pm 200$~K and $\textrm{log g}= \textrm{3.0--4.0}$ using different models \citep[BT-SETTL and DRIFT-PHOENIX,][]{2011ASPC..448...91A, 2008MNRAS.391.1854H, 2009A&A...506.1367W, 2011A&A...529A..44W} and spectra, but which remain broadly consistent with our values. 

We present the first estimates of C/O and [M/H] for both companions. Both objects have C/O values that are compatible with solar value ($\textrm{C/O}_{\odot}=0.55$) within $1\upsigma$. However, they have different behavior in terms of metallicity: \shortgscb\ has a solar metallicity (within 1$\upsigma$) while the one of \shortgqlupb\ appears slightly super-solar ($>$3$\upsigma$). We interpret these abundances in section \ref{sec:abundances}.

The extinction of \shortgqlupb\ is consistent with the value found by \cite[$\textrm{Av}\sim2.3$~mag,][]{2021AJ....162..286S}. It is above the values reported on GQ Lup A and C \citep[$\textrm{Av} \sim 1.0$~mag,][]{mcelwain_first_2007, 2020A&A...635L...1A}. The extinction of \shortgscb\ is within the range of values measured on the primary star \citep[$\textrm{Av} = 0.7 \pm 0.5$~mag,][]{2016ApJ...827..142B, 2022A&A...658A..91A}.

\subsubsection{Use of evolutionary tracks}

Using our estimated \Teff\ from \formosa, recent "hot-start" evolution models \citep{2020A&A...637A..38P} predict a mass of $15.5\pm0.5$~\MJup at the age of the system (Sect.~\ref{subsec:GSC06214b}) for \shortgscb, consistent with previous estimates based on the bollometric luminosity. Our estimated surface gravity is broadly consistent with expectations at the age of UCL ($\textrm{log g}=4.07\pm0.01$) and Upper Scorpius ($\textrm{log g}= 4.04\pm 0.02$) and does not allow for discrimination between the two possible system memberships. Predicted radii range from 1.8 to 2.0~\Rjup, e.g. $\sim20$\% larger than the value found with atmospheric models alone. The mismatch amplitude is consistent with the predictions of the same atmospheric models for VHS1256b, a cooler ($\textrm{T}_{\textrm{eff}}\sim 1400$~K) companion of the same mass range as \shortgscb\ and might point to a systematic overestimation of the absolute near-infrared fluxes by the ATMO models across a broad \Teff range \citep{2023A&A...670L...9P}. This is, however, not an issue for removing the photospheric contribution since the radii are only used to scale the synthetic spectrum to the companion continuum emission level. 

The use of \gqlupb's \Teff from \formosa\ as input for the ATMO evolutionary tracks predict a mass of $22^{+2}_{-3}$~\MJup, a radius in the range $2.65-3.3$~\Rjup, and $\textrm{log g}\sim3.8-3.9$~dex at the system age (Section \ref{subsec:GQLupb}). Both the predicted radius and log~g are inconsistent with those inferred from \formosa\ and the deviation is at odd with that observed for \shortgscb. The use of the older AMES-DUSTY tracks \citep{2000ApJ...542..464C} do not change the conclusions. This could arise from uncertainties in the evolutionary tracks not capturing the physics of the interior of this more massive companion (e.g., deuterium burning). However, the radius is degenerate with the extinction in our fit and higher extinction than estimated would naturally lead to a better agreement between \formosa\ and  evolutionary model predictions.
\cite{2021AJ....162..286S} find radii of $3.55-4.13$~\Rjup\ and a higher extinction (2.3-2.7~mag), which remain slightly inconsistent with the \cite{2020A&A...637A..38P} evolutionary models. The large radii and inconsistent log~g derived from atmospheric models could also be produced because of additional flux contribution such as veiling increasing the flux and requiring a larger dilution factor (Sect.~\ref{sec:discussion:continuum_variability}). A similar mismatch has also been observed on other companions of the same \Teff range in Upper Scorpius \citep{petrus_new_2020} and unresolved binarity has been proposed as an explanation.
 Should \shortgqlupb\ be an equal-luminosity equal \Teff binary, the apparent J-band flux would be deviant overestimated by a factor two with respect to a single object. In such a case, the corresponding radius of individual components found with \formosa\ ($3^{+0.2}_{-0.1}$~\Rjup) would match well the evolutionary model predictions for the fitted \Teff.



\begin{figure*}
    \centering
    \includegraphics[width=\columnwidth]{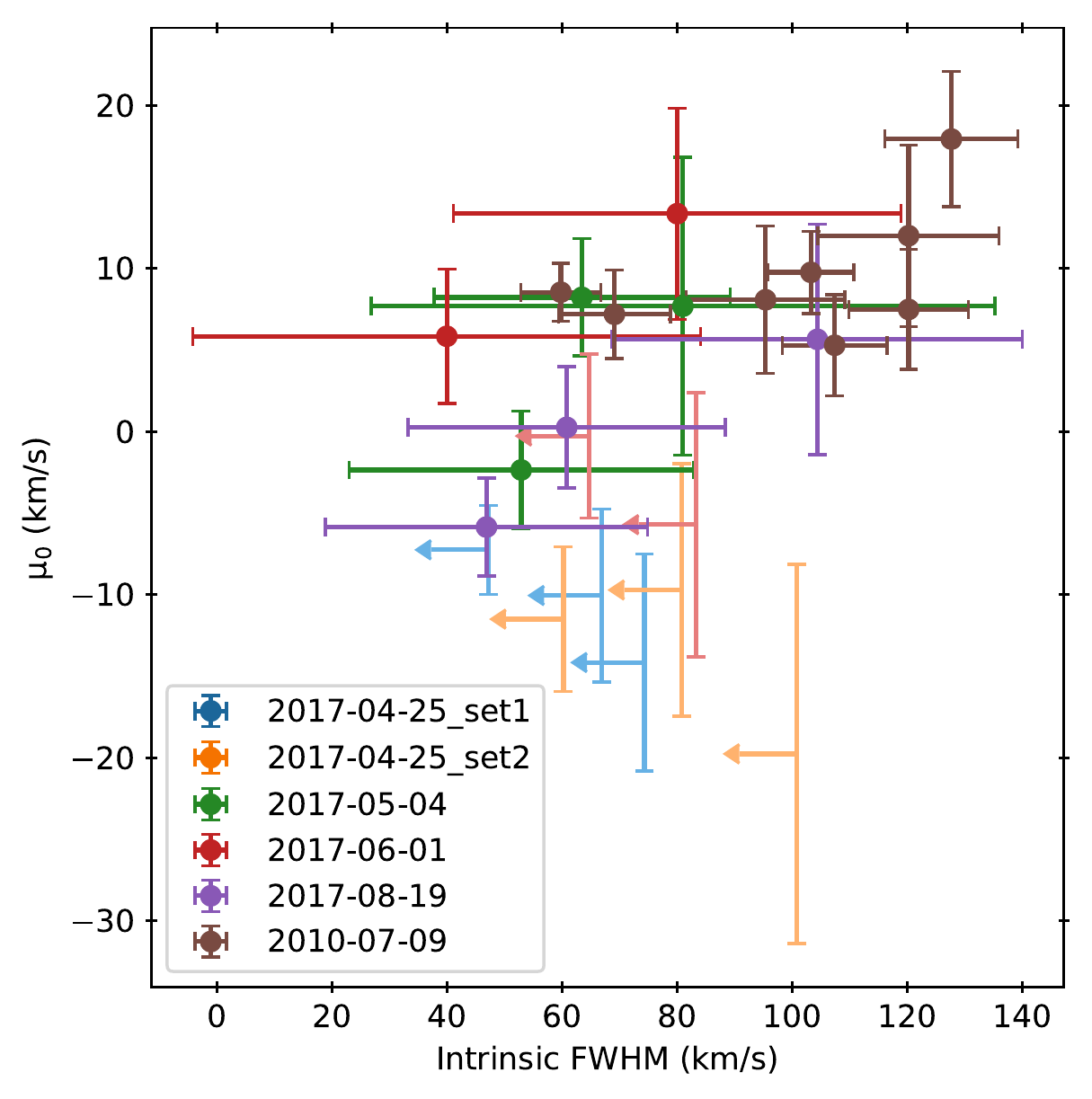}
    \includegraphics[width=\columnwidth]{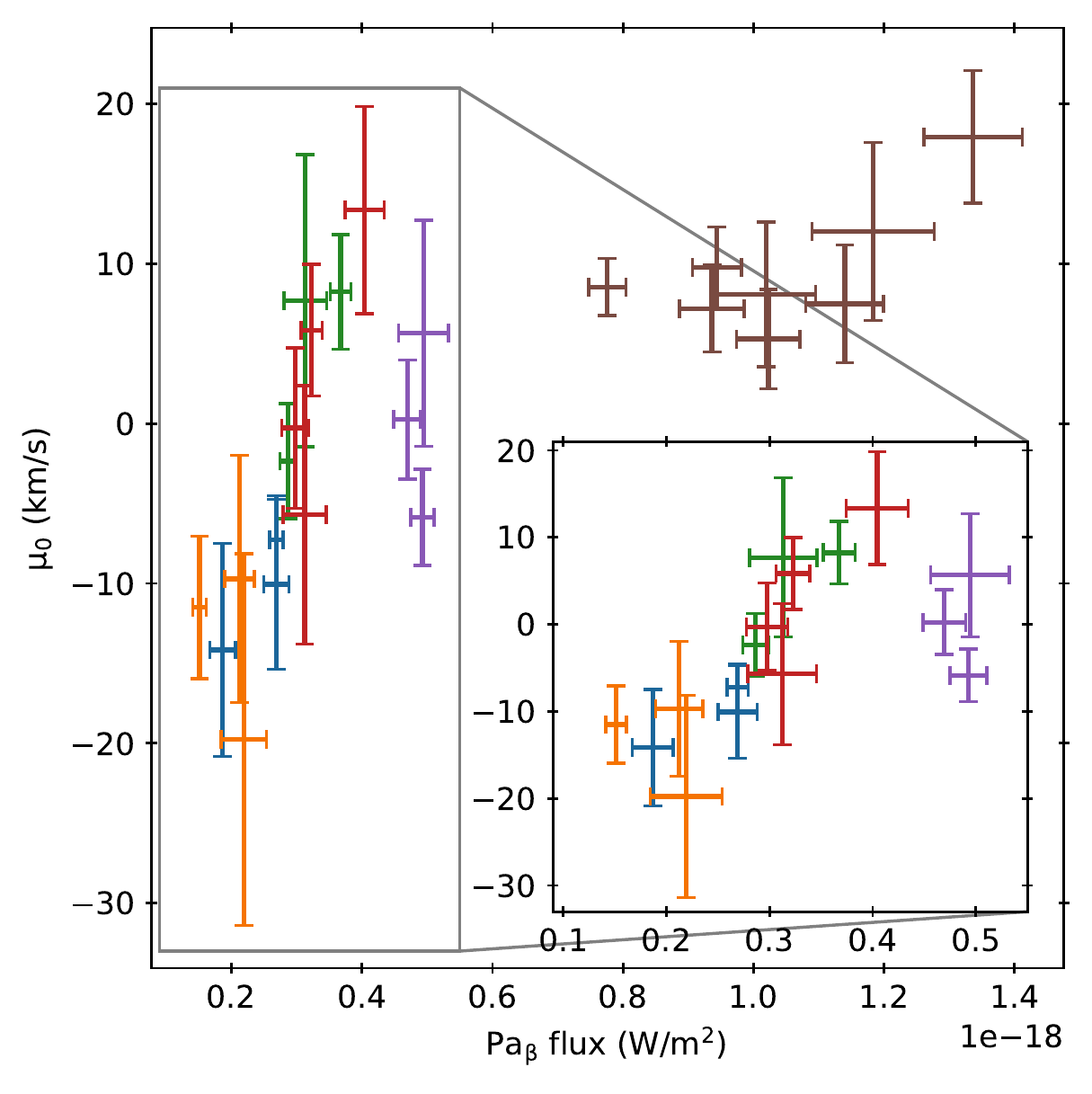}
    \caption{Gaussian-fitting resulting parameters for the \PaBeta\ line of \shortgscb. Left: correlations between intrinsic FWHM and the line peak location $\upmu_0$, arrows represent an upper limit on the intrinsic FWHM. Right: correlations between the \PaBeta\ line flux and the line peak location $\upmu_0$. The intrinsic FWHM is computed as the square root of the quadratic difference between the measured line FWHM and the LSF FWHM. The $\upmu_0$ parameter is relative to the line expected wavelength in vacuum (1.2821578~\um).}
    \label{fig:results:correlations_fwhm_mu0}
\end{figure*}

\subsection{Analysis of \PaBeta\ time series}
\label{sec:results:analysis_pabeta}

\PaBeta\ emission line profiles for individual data cubes are presented in Fig.~\ref{fig:individual_lines}. We performed Gaussian fitting of the lines to infer their width, position and flux (see Table~\ref{tab:results:gaussian_fitting}).

We find the \PaBeta\ line profiles to be partly resolved in the 2006 epoch of \shortgqlupb\ (1--1.5 times the LSF width, 1--3~$\upsigma$), and mostly unresolved in the 2017 epochs of \shortgscb\ (1--1.2 times the LSF width, 1--2~$\upsigma$). Although, we do find the line to be partly resolved in the 2010-07-09 epoch (OSIRIS) (1.2--1.9 the LSF width, 4--8~$\upsigma$). See Table~\ref{tab:results:gaussian_fitting} for details.
We also find the peak of the \PaBeta\ emission lines in \shortgqlupb\ to be slightly blue-shifted. For \shortgscb, they are mostly centered at zero velocity and show no significant asymmetries (with respect to the line center), except for the 2017-06-01 epoch of \shortgscb\ for which the line has a wider red-wing in comparison to the blue wing. This is reminiscent of emission lines formed in magnetospheric accretion columns in T Tauri stars, which show prominent blue/red asymmetry \citep{edwards_spectroscopic_1994}. Shock induced accretion lines also produce lines with a red-shifted peak, hence a wider red wing. These will be explored in more details in Sect.~\ref{sec:discussion:constraints_accretion_processes}.

We studied both the line profiles and fluxes variability. Note that studying integrated fluxes is equivalent to studying EWs, as they were all computed on a similar continuum baseline. Integrated fluxes were computed by Gaussian fitting of the line profiles after local baseline correction (polynomial, n=2), for which main results are reported in Table~\ref{tab:results:gaussian_fitting} and shown in Fig.~\ref{fig:results:correlations_fwhm_mu0}. 

As illustrated in Fig.~\ref{fig:results:correlations_fwhm_mu0}, we find, for \shortgscb, possible correlations between the line intrinsic width and its peak velocity ($\textrm{FWHM}_{\textrm{in}}$--$\upmu_0$, left panel), and between the line flux and its peak velocity (F(\PaBeta)--$\upmu_0$, right panel).
We computed Pearson's correlations coefficients and p-values using an MCMC approach almost equal to the \texttt{pymccorrelation} tool \citep{curran_monte_2015, privon_hard_2020}, with the added constraint that the intrinsic FWHM distributions cannot be negative.
For the first correlation, ignoring upper limits we find $r=0.39^{+0.19}_{-0.22}$ and $\textrm{log p} = -0.88^{+0.59}_{-0.87}$. Note that while they were excluded, upper limits still appear in the correct region for the trend.
For the second correlation (excluding the 2010-07-09 epoch), we find $r=0.52^{+0.10}_{-0.12}$ and $\textrm{log p} = -1.43^{+0.52}_{-0.61}$.
The $\textrm{FWHM}_{\textrm{in}}$--$\upmu_0$ correlation, if true, could be the sign of either (i) red-shifted absorption within the line (as absorption increases, the line centroid will be moved to the blue) or (ii) increasing red-wing contribution as the line flux increases.
The F(\PaBeta)--$\upmu_0$ correlation, if true could be explained in the same way as the (ii) explanation for the peak-velocity/FWHM correlation. Possible implications are discussed in Sect.~\ref{sec:discussion:constraints_accretion_processes}.

From Fig.~\ref{fig:individual_lines}, we see unambiguous EW variability from an epoch to another for both sources. There is no clear sign of line profile variability within the SINFONI sequences -- at the exception of a possible red-wing variability on the \shortgscb\ 2017-06-01 epoch. However, we do see line flux variability within multiple sequences of \shortgscb, in particular the 2010-07-09 one.

\begin{figure*}
    \centering
    \includegraphics[width=\textwidth]{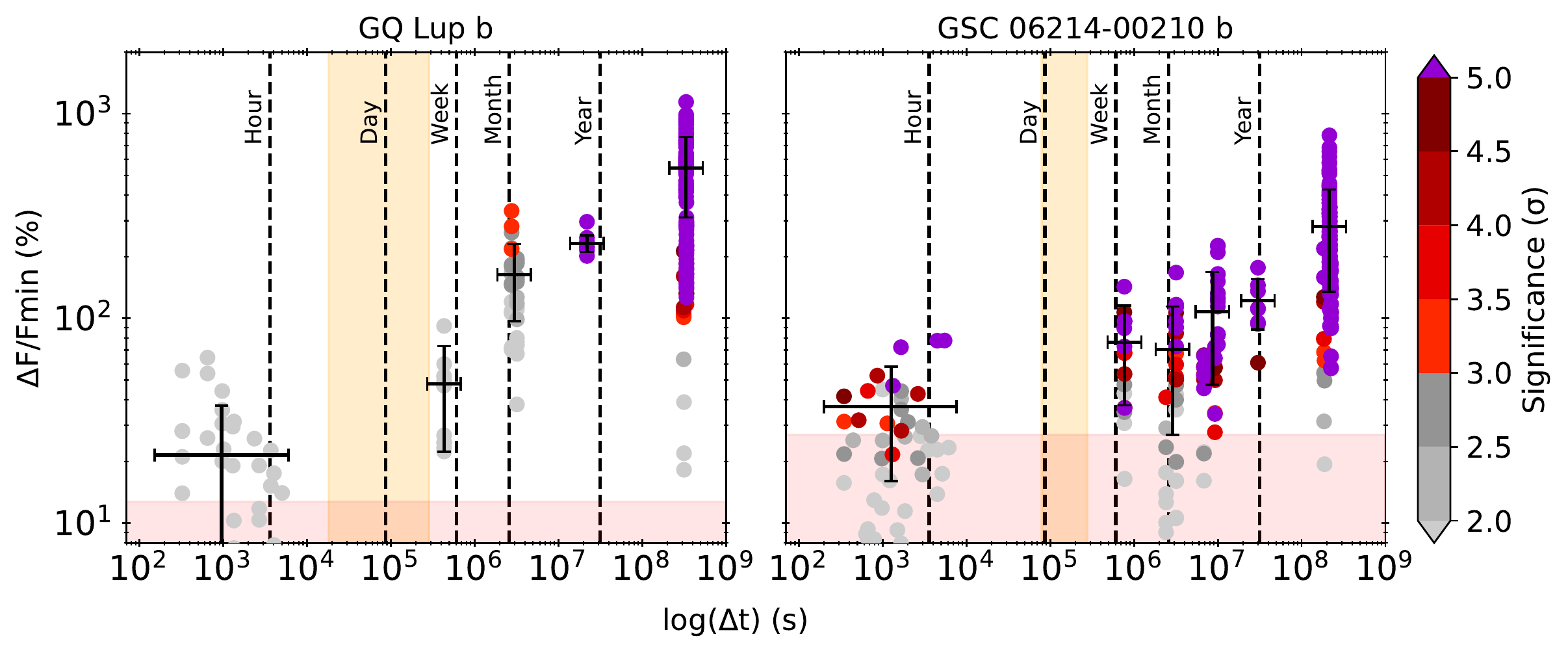}
    \caption{\PaBeta flux variability for both sources, as a function of the timescale, with respect to the lowest-flux point of the pair. For \shortgqlupb, are included data points from this study as well as the spectrum from \cite{lavigne_near-infrared_2009}. For \shortgscb, are included data points from this study as well as those from the individual Bowler data cubes \citep{bowler_disk_2011} and the mean spectrum from \cite{lachapelle_characterization_2015}. The color of the points represents the significance of the measurement, i.e., the $\upsigma$-distance between the two lines used to compute this value. The horizontal and vertical bars correspond respectively to the bin size, and the spread (weighted by significance) within that bin. The red-shaded regions mark the average observed continuum variability among the SINFONI sequences (see Table~\ref{tab:obs}). The orange shaded regions represent the expected rotational period of the objects (see Sect.~\ref{sec:section1}).}
    
    \label{fig:results:variability_timescales}
\end{figure*}

Figure~\ref{fig:results:variability_timescales} shows the relative \PaBeta\ flux variation, computed as (Fmax-Fmin)/Fmin for each individual pair of lines, as a function of their time difference. The marker's color reports the significance of the measurement, i.e., the $\upsigma$-distance between the two lines used to compute this value. While a small and noisy denominator may artificially increase the relative variation, the measurement will appear gray as the two values would not be statistically far apart. In other words, gray measurements are compatible with no relative flux variation (given our uncertainties), while colored points highlight an actual variation (given the $\upsigma$ level).
\shortgqlupb\ shows overall close to no flux variability ($3 \upsigma$ confidence level, $<50$\%) at the hour timescales. However, we do see significant variability ($>100$\%) at the $\sim$month and longer timescales. Note that the $\sim$year timescale corresponds to the variation between the 2006-09-18 and the \cite{lavigne_near-infrared_2009} (2007-05-30) epochs.
\shortgscb, on the other hand, shows definitive flux variability around the 30-min and $\sim$hour timescales ($> 5 \upsigma$ significance). We also do see clear ($>5\upsigma$) line variability at the weeks, months and years timescales. The line flux variability at the $\sim$hour timescale appears within multiple datasets: the OSIRIS epoch (2010-07-09), the 2017-05-04 epoch and between the two sets of the 2017-04-25 epochs. In both objects, the amplitude of variability stays below 50--60\% for timescales smaller than their estimated rotation periods, while stronger variability amplitude (up to a factor 10) is observed at longer timescales. We discuss these findings in the more general context of variable young stars in  Sect.~\ref{sec:discussion:constraints_accretion_processes}.

\begin{figure*}
    \centering
    \includegraphics[width=0.99\textwidth]{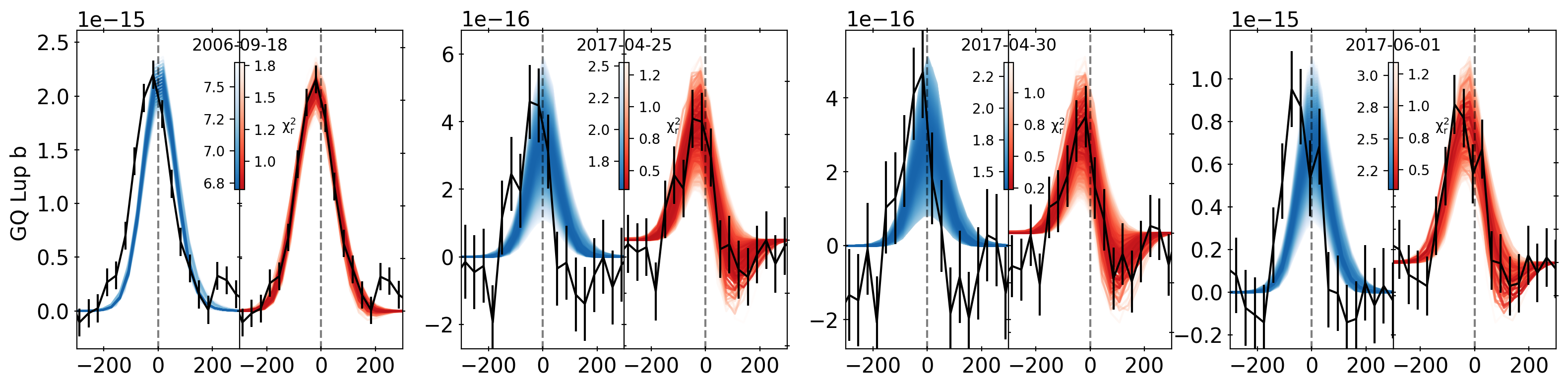}\\
    \includegraphics[width=0.99\textwidth]{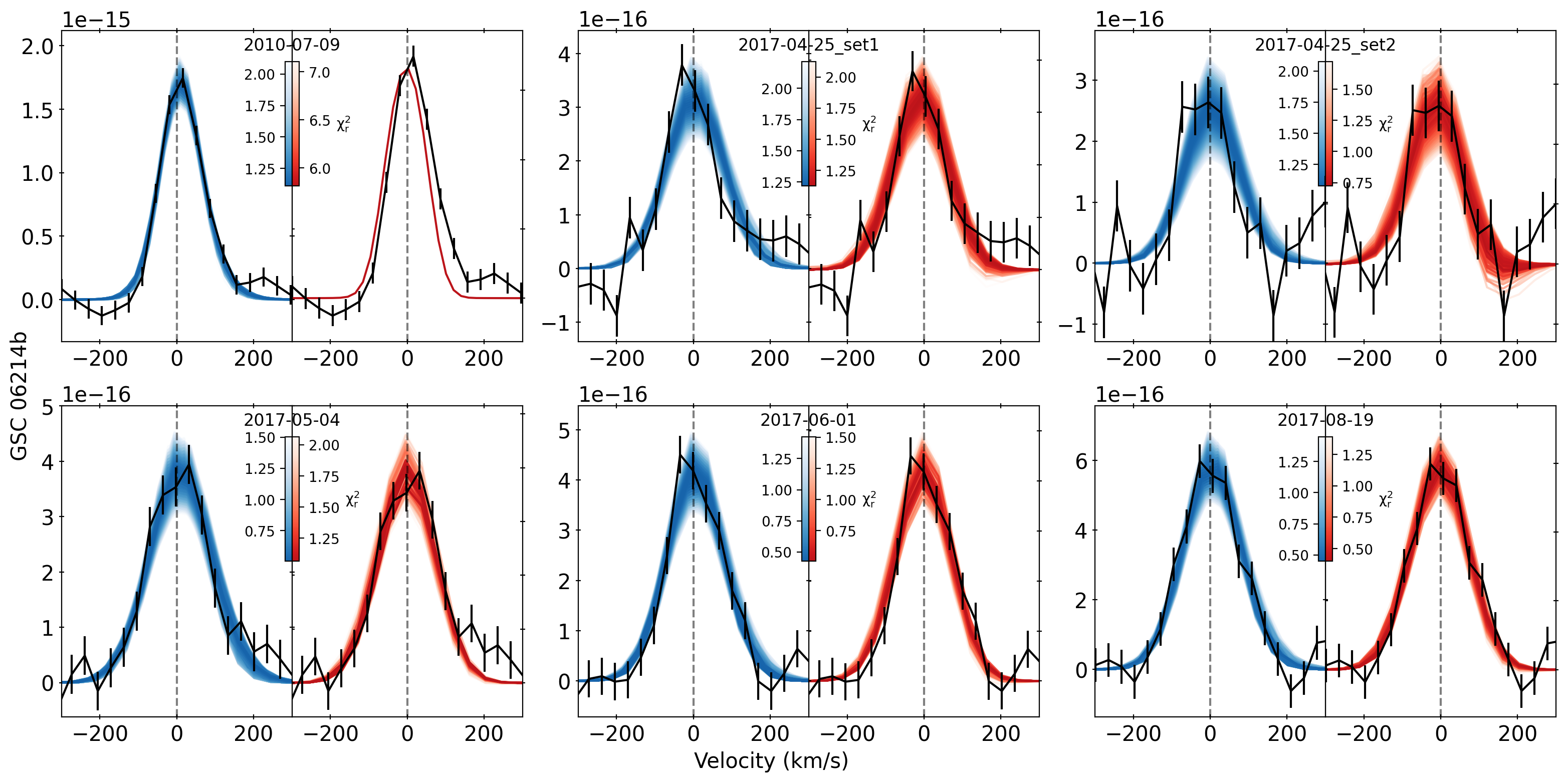}
    
    \caption{Comparison of the best fitting models for both \shortgqlupb\ (first row) and \shortgscb\ (2nd and 3rd rows). Colorbars represent the reduced $\chi^2$ ($\upchi^2_r$) of the Shock (blue) and MagAcc (red) models. For each epoch, all best models (within the $\upchi^2_r$ range given in the colorbar) are over-plotted with a color scaling with the reduced $\chi^2$ color. The fitting limits and number of parameters of each model yield a number of degree of freedom equal to 14 for the Shock model and 9 for the MagAcc model.}
    \label{fig:results:model_fitting_comparison}

    \bigskip
    
    \includegraphics[width=0.499\textwidth]{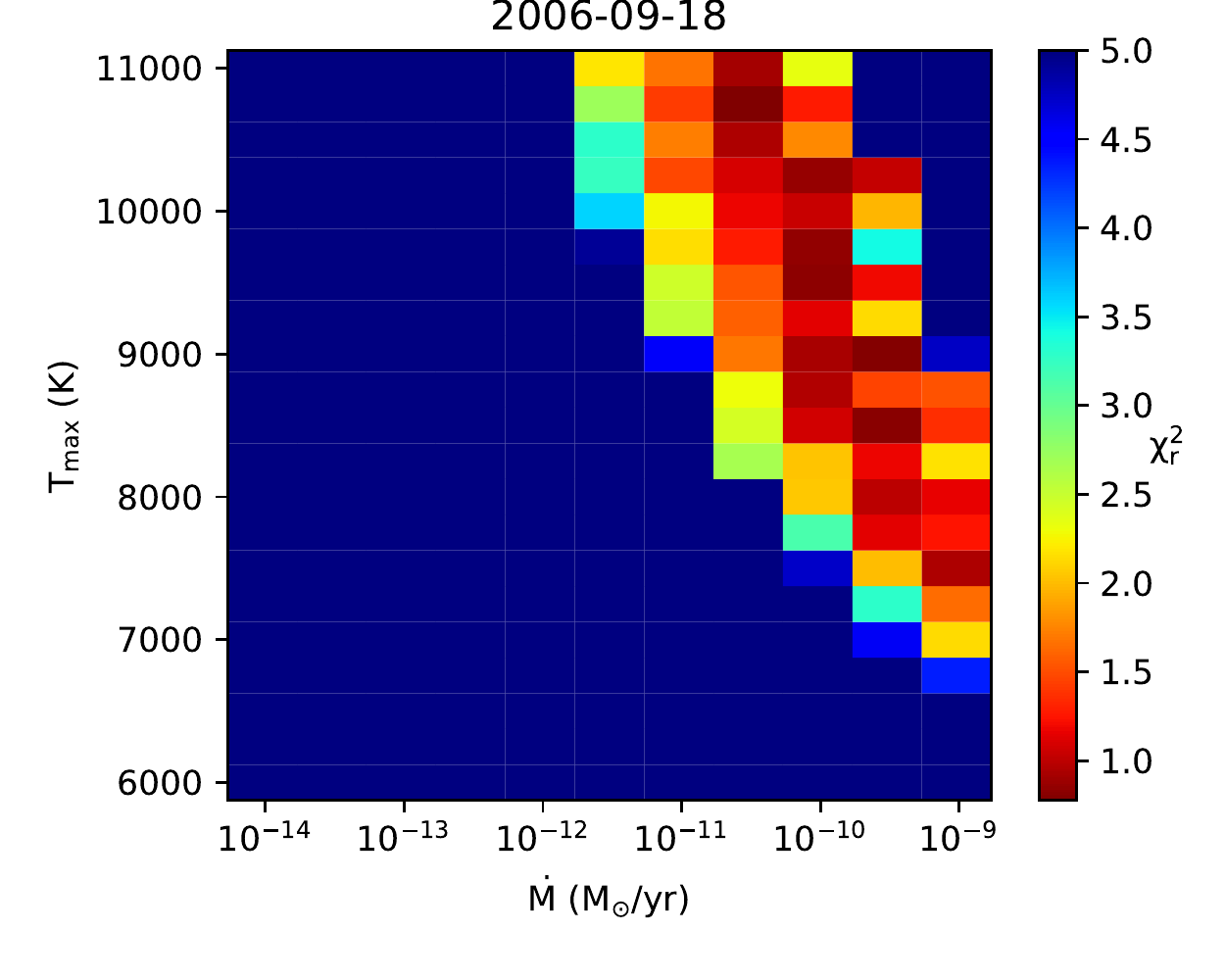}%
    \includegraphics[width=0.499\textwidth]{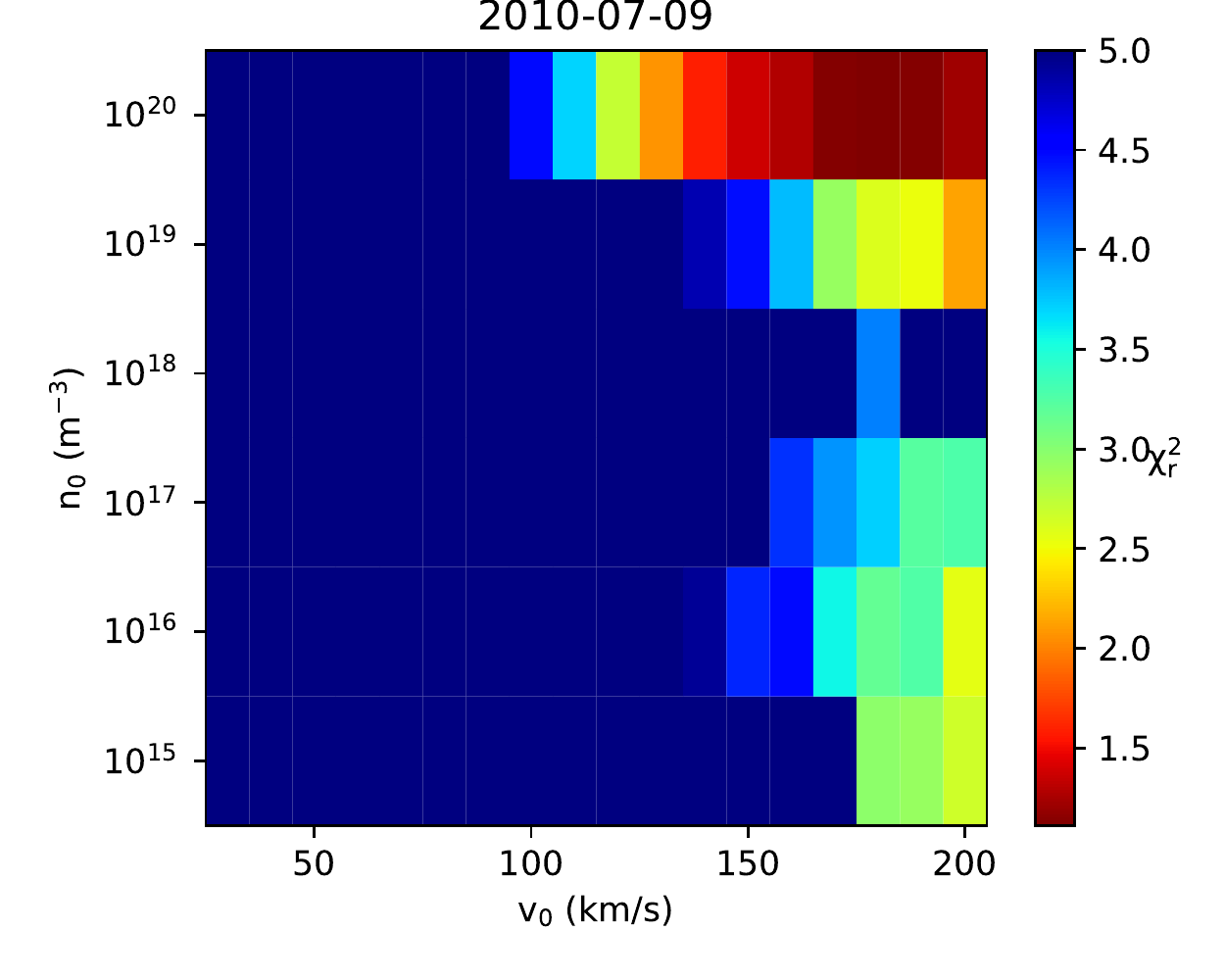}
    \caption{$\upchi^{2}_r$ maps illustrating the accretion models degeneracies when fitting the \PaBeta\ lines. Left panel: \Tmax\ vs \Mdot\ degeneracy for the MagAcc model (\shortgqlupb, epoch: 2006-09-18). Right panel: \nz\ vs \vz\ degeneracy for the Shock model (\shortgscb, epoch: 2010-07-09 -- OSIRIS)}
    \label{fig:results:models_parameters_degenracies}
\end{figure*}

\subsection{Comparison to protoplanet accretion models}

\PaBeta\ emission in the spectra of low-mass free-floating brown-dwarfs of the same mass, age, and spectral-type range as the two companions studied here have been investigated by \cite{2004A&A...424..603N} and \cite{2013A&A...551A.107M}. The \PaBeta\ line intensity is shown to be strongly correlated with the accretion luminosity, while the chromospheric activity does not contribute significantly to the line intensity for accretion rates above $10^{-11}-10^{-12}$ M$_{\odot}$\,yr$^{-1}$ at ages 3--10~Myr. Furthermore, none of the non-accreting sources display \PaBeta\ emission lines. We therefore assume here that most, if not all of the \PaBeta\ emission, is produced by accretion. In this section, we explore two accretion models in the planetary mass regime predicting \PaBeta\ line flux and profiles, and compare their predictions to our time-series observations, in order to infer the corresponding physical parameters.

\subsubsection{Description of accretion models}

\cite{thanathibodee_magnetospheric_2019} proposed a version of the magnetospheric accretion models developed for Classical T Tauri Stars (CTTS) \citep[see the review of][and ref. therein]{hartmann_accretion_2016} extended to the planetary mass range.  It assumes that the central object's dipolar magnetic field is intense enough at young ages to truncate the inner CPD, lift material from the CPD and channel it towards the central object. Material then falls down to the planetary surface at free fall velocity along accretion funnels. In these funnels, the material is heated to temperatures high enough ($\sim 10^4$~K) to trigger atomic Hydrogen line emission (\Halpha, \PaBeta, \BrGamma, ...).
The free-falling material then creates a shock at the base of the funnel flows when it hits the planetary surface, which triggers hot continuum emission that can be absorbed by the infalling gas, leading to red-shifted absorption features within the emission line profiles. Red-shifted absorption also occurs when the accretion flow absorbs the photosphere emission itself.
Blue-shifted absorption may also be present due to jets or outflows, but these are not accounted for by these models.
The magnetospheric accretion model from \cite{thanathibodee_magnetospheric_2019} is described by: the accretion column launching radius \Rin, its \width, its maximum temperature \Tmax, the accretion rate \MdotMagAcc, the \inclination, the planet mass \mass, the planet radius \Rp, and the planet effective temperature \Teff. The model is axis-symmetric and assumes that the large scale dipolar magnetic field is aligned with the central object rotation axis. The model neglects line emission contribution from the shock at the base of the accretion funnels. This model will be referred to as the MagAcc model hereafter.

Conversely, \cite{aoyama_theoretical_2018} proposed that accretion lines, rather than being triggered within accretion funnels, would be triggered within an accretion shock either at the planet or the CPD surface. In such a scenario, material flows from the circumstellar disk down to the planetary or CPD surface.
This may happen for planet embedded within the circumstellar disk of their host star, through polar-inflow \citep{marleau_accreting_2022} or, as was proposed for \shortgqlupb, if material from the circumstellar disk channels up to the planet immediate environment \citep{2021AJ....162..286S}. Alternatively, this model could also describe line emission formed in the shock at the base of magnetospheric accretion columns. 
This model considers two parameters: the infall velocity \vz\ and the pre-shock density \nz. We add an additional flux scaling parameter, \Seff, the effective shock surface (i.e., $4 \pi R_p^2\ffill$ if it takes place at the planetary surface). See details in next section. This model will be referred to as the Shock model hereafter.

\subsubsection{Model fitting} \label{sec:results:model_fitting_results}
We used both aforementioned models \citep[MagAcc and Shock,][]{thanathibodee_magnetospheric_2019,aoyama_theoretical_2018} to try and retrieve physical parameters of the accretion mechanisms by fitting the \PaBeta\ line of both \shortgqlupb\ and \shortgscb.

Both models were scaled to the distance of the objects (Sect.~\ref{sec:section1}). For the MagAcc model, this was performed by applying a factor (\Rp/$\textrm{D})^2$, with D the distance and \Rp\ the planet radius.
For the Shock model, we applied a factor \Seff/(4$\uppi \textrm{D}^2$).
The fits were performed on photosphere-subtracted and baseline-corrected \PaBeta\ emission line profiles, with continuum-subtracted models.
This allows accretion models fitting not to be biased by continuum discrepancies, and to focus solely on the line profile. The models are then convolved to the instruments resolution (R=2360 and R=1800 for \shortgqlupb\ and \shortgscb\ SINFONI data respectively, R=3800 for OSIRIS).

The fits were performed on the mean emission line of each epoch, computing the $\chi^2_r$ on the whole grid of models. Fiducial results are presented in Fig.~\ref{fig:results:model_fitting_comparison}. The \shortgqlupb\ lines are only reproduced by the MagAcc models: this is because the Shock models cannot reproduce a blue shifted emission line peak, as is clearly apparent from the first row of Fig.~\ref{fig:results:model_fitting_comparison}.
Note that the red-shifted absorption in the 2017 epochs of \shortgqlupb\ and fitted by the models is mostly likely caused by noise in the data.
However, \shortgscb\ lines are  well reproduced by both models. Maps of the best $\chi^{2}_r$ are shown in Fig.~\ref{fig:results:models_parameters_degenracies}. We only represent the \Teff--\MdotMagAcc\ axis in the figure for the MagAcc model, because others parameters are  mostly unconstrained.

In the case of the MagAcc scenario, there is a clear degeneracy between the accretion column temperature and the accretion rate. This behavior was already seen in the context of magnetospheric accretion on T Tauri stars \citep{muzerolle_emissionline_2001} and the degeneracy could be lifted with very high resolution observations and/or fitting multiple accretion lines simultaneously.
This shows that in the magnetospheric accretion framework, we cannot derive an accurate measurement of the accretion rate, but only a lower limit.
For \shortgqlupb, this leads to $\Mdot \gtrsim 10^{-11}$ \Msun/yr for the (bright) 2006-09-18 epoch and  $\Mdot \gtrsim 10^{-11.5}$\Msun/yr for the other epochs. 
For \shortgscb, this leads to a lower limit of $\Mdot \gtrsim 10^{-12}$\Msun/yr  for all epochs.
\cite{2014ApJ...783L..17Z} derived values of $10^{-10.8}$ \Msun/yr and $10^{-9.3}$ \Msun/yr from fitting the UV-optical excess emission for \shortgscb\ and \shortgqlupb\ respectively, which are consistent with our lower limits.

\shortgscb\ lines are also well reproduced with the Shock model.
In this scenario, fits are mostly degenerated over the parameter space, except for the 2010-07-09 (OSIRIS) epoch for which high velocity models are strongly favored, as shown in Fig.~\ref{fig:results:models_parameters_degenracies}.
The high shock velocity models correspond to infall velocities $v_{ff} > 150$~\kms, which imply a shock location at $R<2.4$~\Rjup, consistent with the estimated radius of the object (Sect.~\ref{sec:results:photosphere_fitting}). These solutions would indicate that the shock is located at the planetary surface. However, lower infall velocities in the range 30--170~\kms\ cannot be fully excluded. They are consistent with a shock located anywhere within 60~\Rjup, so either at the protoplanet surface or on the CPD at larger scales. In the 2010-07-09 (OSIRIS) epoch, our best fit values of \Seff, \vz\ and \nz\ ($1.5\times 10^{-2}$~S$_{\textrm{jup}}$, 180~\kms and $10^{20}$~$\textrm{m}^{-3}$) yield a mass accretion rate of \Mdot\ $\sim$ $10^{-9}$~\Msun/yr.
Assuming a shock at the planetary surface, the accretion rate can be given as $\Mdot = S \mu \nu_0 v_0$, with $\upmu$ the mean molecular weight \citep[$3.84 \times 10^{-27}$~kg as per][]{aoyama_theoretical_2018}. Inverting the equation, we find the shock surface given as a fraction of the planetary surface (with an estimated radius of 1.8--2.0~\Rjup), this corresponds to a filling factor of 0.38--0.47~\%. This is of the same order (0.3\%) as reported for the older analogue companion Delorme 1~(AB)b  and using similar models on Balmer lines \citep{2023A&A...669L..12R}.

\section{Discussion} \label{sec:discussion}
In this section, we first interpret the derived atmospheric abundances of both objects. We then discuss the implications brought by our results for the origin of the \HI\ line emission in planetary mass objects and the possible impact of  variable veiling. We finally extrapolate how emission line variability as seen here could impact protoplanet detection surveys.

\subsection{Atmospheric abundances as formation tracers}
\label{sec:abundances}
The star forming regions Sco-Cen and Lupus I are expected to have nearly solar metallicity \citep[e.g.,][and ref. therein]{2011AJ....142..180B, 2017A&A...605A..66B, 2021MNRAS.502.1633M} and it has been shown that nearby solar-type stars with $\textrm{[M/H]}=0$ have C/O distributed around the solar value \citep{2018A&A...614A..84S, 2022ApJS..259...45T}. We note however a spectroscopic determination of $\textrm{[M/H]} = -0.06\pm0.01$ and $-0.35\pm0.01$~dex for \shortgsca\ and \gqlupa, respectively \citep{2021AJ....161..114S}. The value found for \gqlupa\ is at odd with expectations and implies that \gqlupb\ is significantly enriched in metals ($\Delta \textrm{[M/H]}=0.61^{+0.05}_{-0.10}$) with respect to its host star while \shortgscb\ metallicity remains compatible within 1$\sigma$ with that of \shortgsca. \cite{2021AJ....161..114S} accounted for veiling affecting the depth of stellar absorption lines of \gqlupa\ in their inversion of the high-resolution optical spectra of the source. They chose in particular the less contaminated and emission-free region 5900–5965~\AA\ for the analysis. However, the \Halpha\ line of \gqlupa\ appears as the most intense of their sample and remaining effects of veiling at considered wavelengths might lead to an underestimation of the [M/H] value for that particular star. 

The metallicity of \shortgscb\ and \shortgqlupb\ are both compatible with the envelope of values predicted for planet population synthesis models based on the core-accretion framework assuming a pristine solar-composition \citep[see Fig. 9 of][]{2021A&A...648A..59P}. Disk-instability models predict a wide range of metallicity values, including solar \citep{2011ApJ...735...30B}. If the companions formed inside the circumstellar disks surrounding their host stars, the nearly-solar C/O implies they should have accreted a substantial fraction of solids. In summary, the derived abundances based on our J-band data do not permit to firmly conclude on the formation scenario of either objects, but an in-situ formation via turbulent fragmentation remains plausible \citep[i.e., fragmentation of the solar nebula into the primary and its companion,][]{padoan1999,padoan2004}.

\subsection{Constraints on accretion processes in planetary mass objects} \label{sec:discussion:constraints_accretion_processes}

It is shown in Sect.~\ref{sec:results:model_fitting_results} that \shortgqlupb\ lines are only reproduced by the MagAcc model, while \shortgscb\ lines are well reproduced by both MagAcc and Shock models.

\subsubsection{MagAcc and Shock models relative contributions}
Although we studied the MagAcc and the Shock scenarios independently, they are not mutually exclusive. The magnetospheric accretion scenario also involves a shock at the base of the accretion columns on the planetary surface, while the Shock model makes no assumption as to where the accretion shock takes place. Therefore, the Shock model could very well explain the line emission at the base of a magnetospherically induced funnel.

The relative contributions of the accretion funnels and the accretion shock to the \PaBeta\ emission line may vary with time. 
The MagAcc induced lines are usually asymmetric with a larger blue wing whereas Shock-induced lines tend to be more symmetric with a slightly stronger red wing. A time-modulation of the MagAcc-Shock contributions could then be the key to explain the correlations observed between the line centroid, its width and flux (Fig.~\ref{fig:results:correlations_fwhm_mu0}). 

Figure~2 of \cite{aoyama_comparison_2021} shows the relative contributions of the funnel flows and the planetary shock \Halpha\ emission, assuming a filling factor of 1--10~\% and planet mass of 6~\Mjup (corresponding to PDS~70~b) depending on the accretion rate. Our higher mass range and different line may quantitatively change the accretion rate threshold at which both emissions become comparable, but this will hardly change the qualitative picture. For the mass accretion rate of \shortgscb\ \citep[$10^{-10.8}$ \Msun/yr, ][]{2014ApJ...783L..17Z}, both MagAcc and Shock contributions are about the same order of magnitude. For the mass accretion rate of \shortgqlupb\ \citep[$10^{-9.3}$ \Msun/yr, ][]{2014ApJ...783L..17Z}, the contribution from the accretion funnels largely dominates, which could explain why only the MagAcc model reproduces the blue-shifted lines of \shortgqlupb.

\subsubsection{Others explanations for line variability}
While we focused the MagAcc and Shock scenarii, other physical mechanisms have been proposed to modulate the accretion lines properties. \cite{marleau_accreting_2022} studied the effects of gas and dust extinction on the \Halpha\ line intensity profile. According to their Fig.~5, for the gas extinction to alter significantly the \Halpha\ flux ($A_R>1$) for an object of $\sim15$~\MJup, the accretion rate must be at least $10^{-5}$ \Mjup/yr, which is far above the accretion rates of \shortgqlupb\ or \shortgscb.
Figures~6, 7 and 8 in \cite{marleau_accreting_2022} show intrinsic \Halpha\ line profiles for the same Shock model as in this work, with and without gas extinction, for various accretion flow geometries, planet masses and accretion rates. Some of these \Halpha\ lines show significant asymmetries (in their intrinsic profile), including an absorption of the red-wing of the line.
If the behavior in the \PaBeta\ line is similar, the correlations observed in Fig.~\ref{fig:results:correlations_fwhm_mu0} could be explained by such gas self-absorption effects. As for dust extinction, they find that it should be negligible for masses above a few \Mjup.

\cite{szulagyi_hydrogen_2020} performed 3D simulations of an accreting protoplanet embedded in a circumstellar disk and studied the lines variation due to the variable shock-front (as the planet orbits around the star) and density variations in the protoplanet vicinity. They find the \PaBeta\ line amplitude variability to be about $\sim 8-41$\%  (depending on the modeling hypothesis) for a 10~\MJup\ planet, at keplerian timescales. The different environments of \shortgscb\ and \gqlupb\ (non-embedded) make this scenario implausible for these two objects and the proposed timescales do not match our observations.  

\subsubsection{Line variability on similar objects}
Variability of PMCs has already been seen at various timescales.
In the \PaBeta\ study of DH~Tau~b by \cite{wolff_upper_2017}, the authors report 4 epochs separated by days and weeks. According to their Table~1, they find the \PaBeta\ line to disappear within $\sim5$ weeks ($3.8\upsigma$) and a tentative variability within a week ($80\pm50$\%, $1.7\upsigma$). \cite{eriksson_strong_2020} conducted a multi-epoch \Ha\ study of Delorme~1~(AB)b. Applying the same methods as in Fig.~\ref{fig:results:variability_timescales} to their Table~B.1 data, we find \Ha\ EW variability spanning $\sim30 \pm 20$\% along the $\sim2$ hours sequence, with no difference between the $<1$~h and $>1$~h timescales. In comparison, we find the variability amplitude to be $\sim 7 \pm 7$\% and $\sim 20 \pm 17$\% for \shortgqlupb\ and \shortgscb\ respectively for timescales $<2$h (note that error bars represent the spread of the variability, not to an uncertainty), compatible with Delorme~1~(AB)b. 

\begin{figure*}[t]
    \centering
    \includegraphics[width=\textwidth]{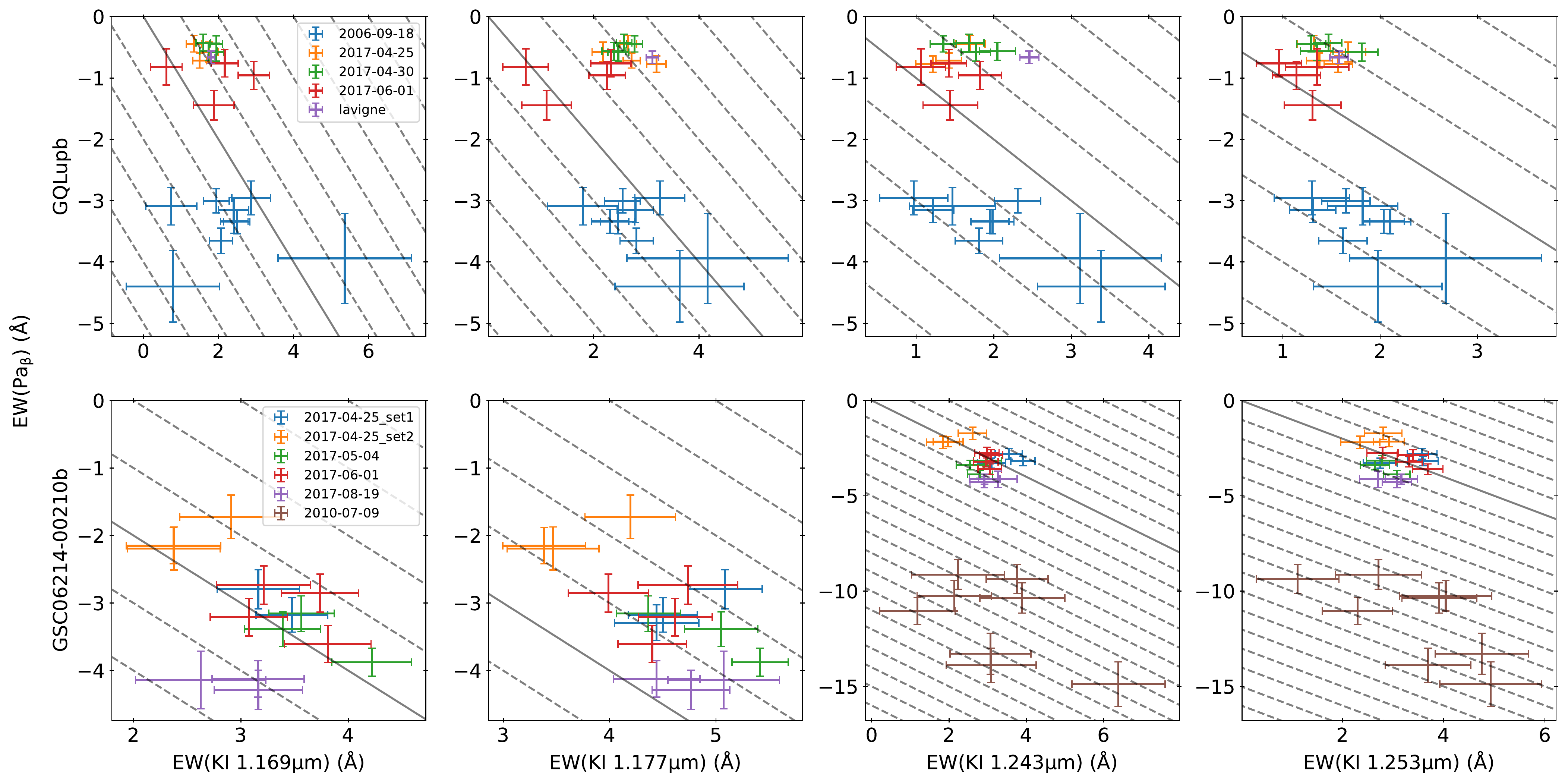}
    \caption{Correlations between equivalent widths (EW) of both potassium absorption doublets and the \PaBeta\ line. The equivalent width is positive for absorption lines and negative for emission lines. Gray lines correspond to a slope equal to -1, where the EW variability can be explained by sole continuum variability.}
    \label{fig:results:KI_correlations}
\end{figure*}

\subsubsection{Similarities with CTTS}
Both \shortgqlupb\ and \shortgscb\ show maximum variability for timescales above their estimated rotation periods (see Fig~\ref{fig:results:variability_timescales}). This behavior is also seen in CTTS \citep{costigan_temperaments_2014, zsidi_accretion_2022}. The \PaBeta\ line profiles of both companions also show strong similarities with the centrally peaked line profiles observed in solar mass accreting young stars \citep[CTTS, see][]{folha_near_2001}.
The line profile of \shortgscb\ on May 04, 2017 and August 19, 2017 matches surprisingly well the one of DG~Tau observed at October 10, 1994 and December 15, 1995 by \cite{folha_near_2001}. The blue-shifted \PaBeta\ line of \shortgqlupb\ is also well reproduced by their observations of DD~Tau. Both objects are sub-solar-mass CTTS.
Magnetospheric accretion models have been very successful at reproducing these line profiles \citep{muzerolle_emissionline_2001}.
These similarities suggest that the mechanisms proposed for the formation of the \HI\ emission lines in CTTS (i.e., magnetospheric accretion) are at play down to the planetary mass regime. The estimates of mass accretion rates for \shortgscb\ and \shortgqlupb\ are also in line with the global decrease of mass accretion rates with object mass which extends down to isolated $\sim$15\MJup\ objects \citep[e.g.,][]{2005ApJ...625..906M, 2013A&A...558L...7J}. 

\cite{2023A&A...669L..12R} report recent high-resolution UV spectra (2 epochs taken within 1h) of the $13\pm5$~\MJup\ companion Delorme 1~(AB)b where they find line blue-shifted profiles indicative of magnetospheric accretion. While the profiles of these lines cannot be directly compared to those of \PaBeta, this further suggests this mechanism could be universal for young 10-30~\MJup\ companions with ages from 3 to 40~Myr. Both  Delorme 1~(AB)b and \shortgscb\ have estimated filling factors of the accretion shock of the order $\sim$1\%. These values are consistent with those found in CTTS \citep[e.g.,][]{2002ApJ...571..378A, 2002ApJ...580..343B, 2022AJ....164..201P} although the methodologies used to derive these factors on stars can be different.

Magnetospheric accretion holds only if the objects interior can generate powerful magnetic fields. Following Eq.~1 from \cite{hartmann_accretion_2016} with the values found from evolutionary tracks (Sect.~\ref{sec:results:photosphere_fitting}, \shortgqlupb: $\textrm{M}=22$~\Mjup, $\textrm{R}=3$~\Rjup, $\Mdot=10^{-9.3}$~\Msun/yr; \shortgscb: $\textrm{M}=15$~\Mjup, $\textrm{R}=2$~\Rjup, $\Mdot=10^{-10.8}$~\Msun/yr), the lower limits on their rotational period (5h for \shortgqlupb, 20h for \shortgscb) and $\upxi=0.7$, they would require a minimum dipolar magnetic field of $\sim70$~G and $\sim150$~G respectively. These increase to $\sim750$~G and $1.7$~kG for the rotational period upper limits (80h). Magnetic field strengths on young companions are just starting to be investigated \citep{katarzy16,2022AJ....163...15C} but no measurement exists yet.  Recent radio and \Halpha\ observations of low-mass brown-dwarfs and planetary-mass objects in the same mass range as our companions suggest that the magnetic field strength could reach several k-Gauss at early ages \citep{2018ApJS..237...25K}. Time-series of line profiles at higher resolutions (R$>$10000) should further help to place constraints on the MagAcc and Shock models \citep{marleau_accreting_2022}. More direct measurements of the magnetic fields strength on accreting free-floating planets or companions on wider-orbits with spectro-polarimeters (e.g., SPIRou) might be possible and would also help to confirm the MagAcc scenario.

\subsection{Evidence for continuum variability?} \label{sec:discussion:continuum_variability}

In our analysis of the \PaBeta\ line variability, we have assumed that the continuum emission remained constant.
Photospheric emission of young dwarfs  at the M-L transition is not expected to show high-amplitude modulation due to non-homogeneous cloud coverage \citep[e.g., ][]{2018haex.bookE..94A, 2022ApJ...924...68V}.  It remains possible that variable veiling or extinction due to accretion and the circum-"planetary" environment or magnetic spots affect the J-band and could change the contrast between the lines and their surrounding continua, thus producing at least a portion the observed variability. \cite{2002AJ....124.1001C} find a 0.15mag variability at J-band of the  faintest members of Chameleon whose colors (J-K$\sim$0.8-1.0) are close to the one of \shortgqlupb\ and correspond to spectral type in the mid- to late-M range \citep{2022ApJ...924...84C}.
 
 \cite{2014ApJ...783L..17Z} measured the shock contribution from 1D slab models (continuum excess produced by the accretion shock) and the \Halpha\ luminosity of both companions using HST data. Their model indicate a negligible contribution of the slab for \shortgscb\ beyond 1~\um\ and a contamination at the percent level for \shortgqlupb\ at J-band. However, the shock contribution is expected to be linked to the line intensity, this could work as a variable continuum component at J band that could explain observed \PaBeta\ equivalent width variations. Photometric variability of \gqlupb\ is known to be negligible at K-band \citep[$\upsigma_{\textrm{K}}=0.073$~mag,][]{2008A&A...484..281N}, but no monitoring has been conducted at J-band where the effect of veiling could be more important. 

Veiling would both decrease the prominent \KI\ doublets and the \PaBeta\ emission lines. Figure~\ref{fig:results:KI_correlations} shows the EW of \PaBeta\ and the KI absorption lines (corresponding values are given in Table~\ref{tab:results:equivalent_widths}). A variation along a line with slope -1 is expected if the observed EW variability is only due to a variation in the continuum level. \shortgqlupb\ lines show no clear correlations in 2017, although there might be a slight trend in 2006. Because the 2006 epoch and the 2017 epochs are not following the same slope, their EW difference cannot be explained only by continuum variability. To explain the EW variation observed within the 2006 epoch, a variation of the continuum level of $\sim30-50$~\% within 2 hours would be required, corresponding to changes of J-band magnitudes of $0.3-0.4$~mag, compatible with the dispersion of values reported in Table \ref{tab:obs}.

For \shortgscb, the trend appears between all epochs except 2017-08-19 and 2010-07-09. The \PaBeta\ EW variability between those might therefore be explained by continuum variability. Although, in a similar fashion as for \shortgqlupb, the \PaBeta\ EW variation between the 2017-08-19/2010-07-09 and other 2017 epochs cannot be explained only be continuum variability.

Given the current constraints, we assume that both objects have EW variability that is driven by intrinsic line flux variability. Accurate photometric monitoring of both companions at J-band simultaneous to \PaBeta\ emission line variability campaign would be required to fully confirm this hypothesis.

\subsection{Implications for protoplanet detection}
\label{sec:discussion:protoplanets_detection}

Several surveys have been conducted to detect accreting planets nested in circumstellar disks using narrow-band imaging at \Halpha\ 
\citep{2018A&A...618L...9M, 2018ApJ...863L...8W, 2019A&A...622A.156C,2020JATIS...6d5004U, 2020A&A...633A.119Z, 2022A&A...668A.138H, 2022arXiv221102109F}, or integral field spectroscopy targeting  \Halpha\  \citep{2019NatAs...3..749H, 2020A&A...644A.149X} and \PaBeta\ emissions \citep{2017AJ....154...90U, 2021AJ....162..214U}. Thus far, \Halpha\ point source detections have been claimed in  four systems \cite[LkCa15, AB Aur, PDS70, CS Cha;][]{2015Natur.527..342S, 2018A&A...613L...5H, 2022NatAs...6..751C, 2021AJ....161..244Z, 2022arXiv221102109F}. The nature of the detection remains unclear in two of these systems \citep[LkCa15, AB Aur;][]{2015A&A...579A..48W, 2022arXiv221102109F, 2022ApJ...934L..13Z}. With the exception of Delorme~1~(AB)b, which does not seem to be in a protoplanetary-disk, no detection has ever been achieved at \PaBeta. 

PDS~70~b and c are known to be affected by significant in-line extinction \citep{2020AJ....159..222H, 2021AJ....161..148W} caused by surrounding material while they are otherwise found within a large cavity. Extinction is therefore a good candidate to explain the poor yield of \Halpha\ and \PaBeta\ deep imaging searches, especially in disks with significant amount of gas and dust. However, most are single-epoch observations and emission line variability  might also provide an additional explanation for the low yield of these observations. 

\cite{2020A&A...644A.149X} report in their Fig. 11 the \Halpha\ detection limit achieved for 30min of effective VLT/MUSE observations of PDS~70. Their observations of PDS~70 are the deepest presented in the paper and probe deeper apparent line fluxes than all the  narrow-band observations presented in \cite{2019A&A...622A.156C} and \cite{2022A&A...668A.138H}. They also report the measured apparent \Halpha\ emission line flux of the two accreting planets.

Both PDS~70~b~\&~c, unlike \shortgqlupb\ and \shortgscb , are embedded and interacting with the circumstellar disk. It is therefore reasonable to assume that the majority of variability happens at keplerian timescales. Given the variety of ages and environments of known accreting objects, our \PaBeta\ variability analysis might only serve as a first order approximation of the general \Halpha\ variability.
Should our observed \PaBeta\ variations still be representative of \Halpha\ variability of others companions, the 100 to 400\% variability amplitude at year to decade timescales measured for \shortgscb\ and \gqlupb\ would imply that at other epochs, both PDS~70 planets might have been emitting below the VLT/MUSE detection threshold. The various \Halpha\ flux measurements of PDS70~b shown in \cite{2021AJ....161..244Z} tend to indicate that such amplitude of variability is plausible for the companion at the year timescale. Our findings clearly advocate new visits of these young systems to maximize the detection probability.
Similarly, variability might have also helped detection surveys.

\section{Conclusions}
\label{sec:conclusions}

We presented the results of a J-band spectroscopic monitoring campaign of \shortgqlupb\ and \shortgscb\ at medium resolving power ($\textrm{R} = 1800-2360$) with VLT/SINFONI. We re-detect the \PaBeta\ emission line at each epoch. We report below our main conclusions:
\begin{itemize}
    \item We confirm that \shortgqlupb\ is a $\sim2640$~K, $22^{+2}_{-3}$~\Mjup\ companion, and \shortgscb\ a $\sim2250$~K, $15.5 \pm 0.5$~\Mjup\ companion, given photosphere fitting and evolutionary tracks. Note that error bars may not be representative of systematic uncertainties of the models.
    
    \item We measure C/O values on both objects compatible with solar value (\shortgqlupb: C/O = $0.44^{+0.13}_{-0.11}$, \shortgscb: C/O = $0.48^{+0.16}_{-0.12}$, C/O$_{\odot}=0.55$). We find \shortgscb's [M/H] to be compatible with solar value ($-0.16 \pm 0.17$), while \shortgqlupb's is slightly super-solar ($0.23 \pm 0.06$). The uncertainties on the atmospheric parameters are driven by small systematic errors in the data revealed by our multi-epoch observations, while Bayesian inference at single epoch provide too optimistic estimates of the error bars.
        
    \item For both objects, the \PaBeta\ intensity variability is found to be moderate ($<50$~\%) at timescales below their rotation period, and more important at longer timescales (up to $\sim1000$~\% for a decade), a behavior similar to that of CTTS. We studied whether variable veiling could change the contrast of the line above the continuum. We find that such flux component does not explain the long-term variability of either object but we cannot exclude it might explain the shorter term evolution of the lines.
    
    \item Both companions had their maximum recorded line brightness at first (historical) epochs. Their variability amplitude can be up to $\sim1000$~\% at the decade timescale. Should that amplitude extend to \Halpha, this might explain the low yield of \Halpha\ and \PaBeta\ (spectro)-imaging campaigns for protoplanet detection.
    
    \item The blue-shifted \PaBeta\ lines  of \shortgqlupb\ are reproduced only by the protoplanet magnetospheric accretion models. Those of \shortgscb\ are well reproduced by both magnetospheric accretion and shock models at quiescence, while the shock model is favored when the line is more intense (2010 epoch). We find filling factors of the accretion shock below 1\% at that epoch and compatible with the value found on the older companion Delorme 1~(AB)b.

    \item We identify a tentative correlation between \shortgscb's emission line center and its width, which could be explained either by asymmetries in magnetospheric accretion-induced lines, or by gas extinction in shock-induced lines.
\end{itemize}

Our study presents the first monitoring of emission lines of young accreting companions at the edge of the deuterium-burning boundary. It illustrates how variability timescales could be used to investigate whether the phenomena at play are a down-sizing of those happening on protostars or more specific ones. Our results advocate multiple visits of young systems on a year-span at least to maximize the detection probability of young protoplanets. Observations at higher resolution ($\textrm{R}>10000$, e.g. ERIS) would now be required to better constrain the line profiles and emitting mechanism.

\begin{acknowledgements}
We thank the ESO staff for the help during the preparation of the observations and their execution. We are grateful to B. Tessore and J. Bouvier for useful discussions on their magnetospheric accretion models. We acknowledge support in France from the French National Research Agency (ANR) through project grant ANR-20-CE31-0012 and the Programmes Nationaux de Planetologie et de Physique Stellaire (PNP and PNPS). This project has received funding from the European Research Council (ERC) under the European Union’s Horizon 2020 research and innovation programme (COBREX; grant agreement 885 593).

G.-D. M. acknowledges the support of the DFG priority program SPP 1992 ``Exploring the Diversity of Extrasolar Planets'' (MA~9185/1),
and
from the Swiss National Science Foundation under grant
200021\_204847
``PlanetsInTime''.
Parts of this work have been carried out within the framework of the NCCR PlanetS supported by the Swiss National Science Foundation.

This publication makes use of VOSA, developed under the Spanish Virtual Observatory (\url{https://svo.cab.inta-csic.es}) project funded by MCIN/AEI/10.13039/501100011033/ through grant PID2020-112949GB-I00. VOSA has been partially updated by using funding from the European Union's Horizon 2020 Research and Innovation Programme, under Grant Agreement nr.~776403 (EXOPLANETS-A).

\end{acknowledgements}

%
%

\bibliographystyle{aa}
\bibliography{ref}

\clearpage
\begin{appendix}

\section{Supplementary material}

\begin{figure}[h]
\begin{minipage}{\textwidth}
\centering
\includegraphics[width=\textwidth]{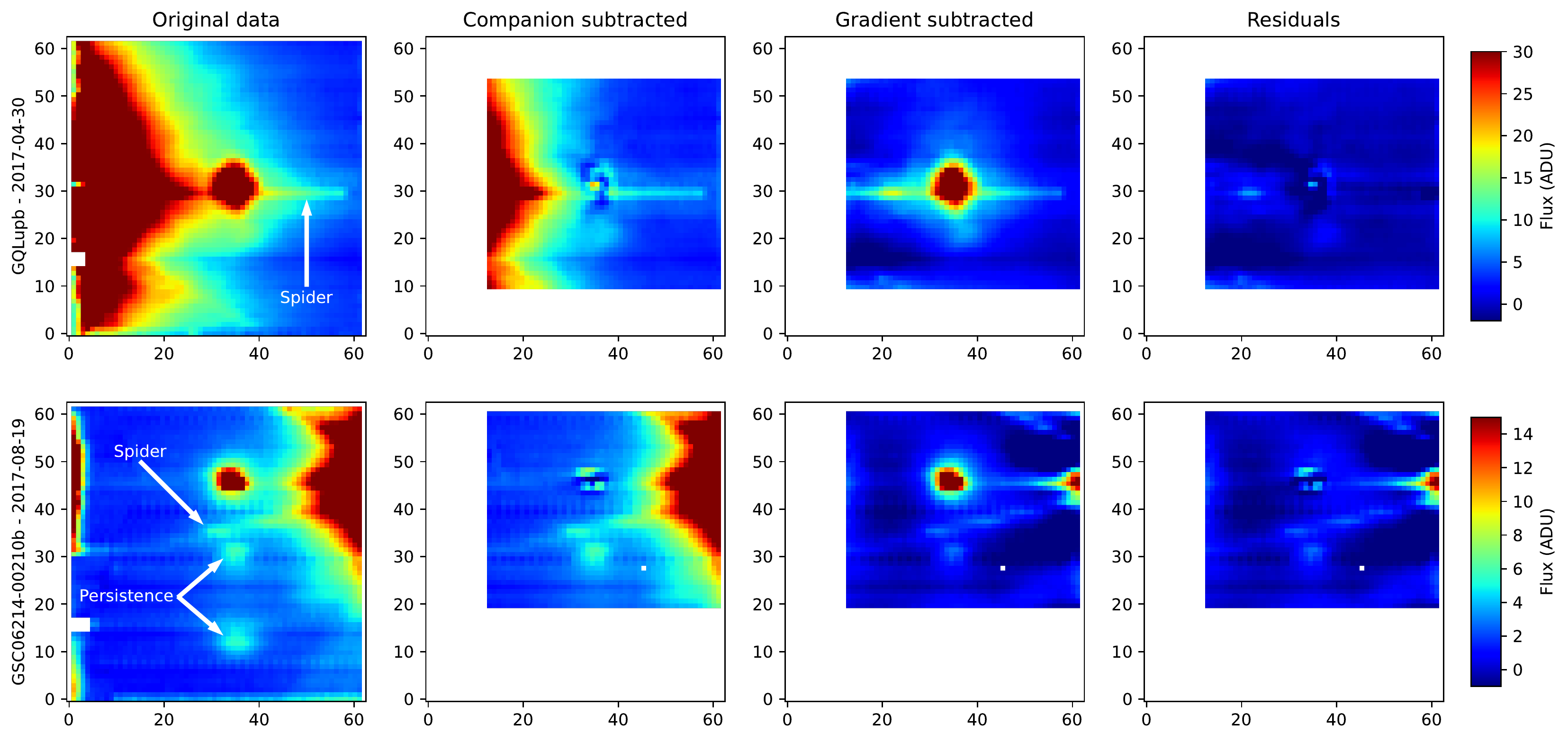}
\caption{Visual representation of the last iteration of the extraction for both \shortgqlupb\ and \shortgscb, on epochs 2017-04-30 and 2017-08-19 respectively. As mentioned in Sect.~\ref{sec:extraction}, the horizontal spider is subtracted for \shortgqlupb, but not for \shortgscb\ as it does not extend below the companion. Are also marked the spiders locations for both targets and some persistence effects in the \shortgscb\ data.}
\label{fig:appendix:residuals}
\end{minipage}
\end{figure}

Figure~\ref{fig:appendix:residuals} contains a visual representation of the extraction process for both objects, along with residuals. Table~\ref{tab:results:equivalent_widths} contains the measured equivalent widths of the two KI doublets and the \PaBeta\ line of each individual cube.
Table~\ref{tab:results:gaussian_fitting} contains the resulting fitted Gaussian parameters of the \PaBeta\ line. The \PaBeta\ integrated fluxes used for Fig.~\ref{fig:results:variability_timescales} were computed from these parameters as $I_{\textrm{tot}} = \sigma I_0 \sqrt{2\pi}$. The error bar on the integrated flux accounts for correlations between parameters:
\begin{equation}
    \left(\frac{\textrm{err}(I_{\textrm{tot}})}{I_{\textrm{tot}}}\right)^2 =  \left(\frac{\textrm{err}(\sigma)}{\sigma}\right)^2 + \left(\frac{\textrm{err}(I_0)}{I_0}\right)^2 + \frac{2\,\textrm{cov}(\sigma, I_0)}{\sigma I_0}.
\end{equation}

\begin{table*}
\caption{Equivalent Widths}             
\label{tab:results:equivalent_widths}      
\begin{center}
\footnotesize
\begin{tabular}{l c c c c c c}
\hline\hline

Source & Date & KI 1.169~\um & KI 1.177~\um & KI 1.243~\um & KI 1.253~\um & \PaBeta\\ 
&  & \AA & \AA & \AA & \AA & \AA\\ 
\hline
\shortgqlupb &2006-09-17 -- 23:22:53 & $2.07 \pm 0.31$ & $2.81 \pm 0.32$ & $1.81 \pm 0.30$ & $1.62 \pm 0.25$ & $-3.65 \pm 0.21$\\ 
\shortgqlupb &2006-09-17 -- 23:39:49 & $2.42 \pm 0.37$ & $2.31 \pm 0.36$ & $1.95 \pm 0.25$ & $2.04 \pm 0.21$ & $-3.34 \pm 0.19$\\ 
\shortgqlupb &2006-09-17 -- 23:45:21 & $2.40 \pm 0.40$ & $2.79 \pm 0.35$ & $1.22 \pm 0.27$ & $1.31 \pm 0.24$ & $-3.15 \pm 0.20$\\ 
\shortgqlupb &2006-09-18 -- 00:02:15 & $2.48 \pm 0.35$ & $2.46 \pm 0.33$ & $1.99 \pm 0.28$ & $2.10 \pm 0.21$ & $-3.34 \pm 0.20$\\ 
\shortgqlupb &2006-09-18 -- 00:07:41 & $1.94 \pm 0.34$ & $2.55 \pm 0.34$ & $2.31 \pm 0.30$ & $1.65 \pm 0.25$ & $-3.00 \pm 0.20$\\ 
\shortgqlupb &2006-09-18 -- 00:24:36 & $2.87 \pm 0.51$ & $3.26 \pm 0.47$ & $0.97 \pm 0.44$ & $1.30 \pm 0.39$ & $-2.96 \pm 0.28$\\ 
\shortgqlupb &2006-09-18 -- 00:30:08 & $0.74 \pm 0.68$ & $1.80 \pm 0.67$ & $1.47 \pm 0.55$ & $1.82 \pm 0.36$ & $-3.09 \pm 0.31$\\ 
\shortgqlupb &2006-09-18 -- 00:47:01 & $0.78 \pm 1.25$ & $3.63 \pm 1.22$ & $3.39 \pm 0.82$ & $1.97 \pm 0.66$ & $-4.40 \pm 0.58$\\ 
\shortgqlupb &2006-09-18 -- 00:52:33 & $5.36 \pm 1.78$ & $4.16 \pm 1.53$ & $3.12 \pm 1.04$ & $2.67 \pm 0.99$ & $-3.94 \pm 0.73$\\ 
\shortgqlupb &2017-04-25 -- 05:21:16 & $1.74 \pm 0.24$ & $2.18 \pm 0.21$ & cosmic & $1.67 \pm 0.18$ & $-0.58 \pm 0.16$\\ 
\shortgqlupb &2017-04-25 -- 05:26:40 & cosmic & $3.19 \pm 0.18$ & $1.21 \pm 0.22$ & $1.57 \pm 0.15$ & $-0.77 \pm 0.13$\\ 
\shortgqlupb &2017-04-25 -- 05:37:30 & $1.34 \pm 0.20$ & $2.66 \pm 0.16$ & $1.70 \pm 0.19$ & $1.32 \pm 0.17$ & $-0.44 \pm 0.13$\\ 
\shortgqlupb &2017-04-25 -- 05:42:55 & $1.50 \pm 0.19$ & $2.72 \pm 0.16$ & $1.42 \pm 0.16$ & $1.38 \pm 0.13$ & $-0.71 \pm 0.12$\\ 
\shortgqlupb &2017-04-30 -- 05:34:35 & $1.94 \pm 0.17$ & $2.78 \pm 0.15$ & $1.35 \pm 0.17$ & $1.29 \pm 0.15$ & $-0.44 \pm 0.13$\\ 
\shortgqlupb &2017-04-30 -- 05:56:16 & $1.94 \pm 0.20$ & $2.47 \pm 0.20$ & $1.77 \pm 0.19$ & $1.81 \pm 0.17$ & $-0.58 \pm 0.15$\\ 
\shortgqlupb &2017-04-30 -- 05:50:51 & $1.73 \pm 0.23$ & $2.39 \pm 0.23$ & $2.05 \pm 0.23$ & $1.35 \pm 0.17$ & $-0.56 \pm 0.15$\\ 
\shortgqlupb &2017-04-30 -- 05:40:00 & $1.59 \pm 0.19$ & $2.58 \pm 0.15$ & $1.68 \pm 0.20$ & $1.47 \pm 0.14$ & $-0.43 \pm 0.14$\\ 
\shortgqlupb &2017-06-01 -- 06:29:29 & $2.16 \pm 0.36$ & $2.33 \pm 0.38$ & $1.42 \pm 0.23$ & $0.96 \pm 0.23$ & $-0.76 \pm 0.22$\\ 
\shortgqlupb &2017-06-01 -- 06:34:55 & $2.94 \pm 0.42$ & $2.25 \pm 0.35$ & $1.82 \pm 0.28$ & $1.14 \pm 0.25$ & $-0.96 \pm 0.23$\\ 
\shortgqlupb &2017-06-01 -- 06:45:44 & $0.61 \pm 0.42$ & $0.71 \pm 0.43$ & $1.06 \pm 0.32$ & $1.35 \pm 0.33$ & $-0.82 \pm 0.29$\\ 
\shortgqlupb &2017-06-01 -- 06:51:09 & $1.87 \pm 0.54$ & $1.11 \pm 0.47$ & $1.44 \pm 0.35$ & $1.31 \pm 0.29$ & $-1.44 \pm 0.24$\\ 
\hline 
\shortgscb &2017-04-25 -- 06:21:07 & $3.16 \pm 0.38$ & $5.09 \pm 0.35$ & $3.55 \pm 0.35$ & $3.58 \pm 0.29$ & $-2.79 \pm 0.29$\\ 
\shortgscb &2017-04-25 -- 06:32:12 & $3.48 \pm 0.33$ & $4.50 \pm 0.33$ & $3.91 \pm 0.31$ & $3.59 \pm 0.30$ & $-3.18 \pm 0.26$\\ 
\shortgscb &2017-04-25 -- 06:48:51 & cosmic & $4.44 \pm 0.40$ & $3.10 \pm 0.36$ & $2.75 \pm 0.33$ & $-3.29 \pm 0.27$\\ 
\shortgscb &2017-04-25 -- 07:35:50 & $2.37 \pm 0.44$ & $3.38 \pm 0.39$ & $1.98 \pm 0.39$ & $2.92 \pm 0.31$ & $-2.15 \pm 0.27$\\ 
\shortgscb &2017-04-25 -- 07:46:53 & $2.37 \pm 0.43$ & $3.47 \pm 0.43$ & $1.84 \pm 0.43$ & $2.35 \pm 0.39$ & $-2.19 \pm 0.32$\\ 
\shortgscb &2017-04-25 -- 08:03:31 & $2.91 \pm 0.48$ & $4.19 \pm 0.43$ & $2.61 \pm 0.37$ & $2.81 \pm 0.37$ & $-1.72 \pm 0.32$\\ 
\shortgscb &2017-05-04 -- 04:29:49 & $3.39 \pm 0.35$ & $5.05 \pm 0.34$ & $2.55 \pm 0.36$ & $2.64 \pm 0.29$ & $-3.38 \pm 0.26$\\ 
\shortgscb &2017-05-04 -- 04:40:54 & $3.56 \pm 0.30$ & $4.37 \pm 0.30$ & $2.99 \pm 0.36$ & $2.78 \pm 0.30$ & $-3.16 \pm 0.26$\\ 
\shortgscb &2017-05-04 -- 04:57:38 & $4.22 \pm 0.37$ & $5.42 \pm 0.26$ & $2.80 \pm 0.33$ & $3.07 \pm 0.26$ & $-3.88 \pm 0.21$\\ 
\shortgscb &2017-06-01 -- 01:54:27 & $3.21 \pm 0.44$ & $4.74 \pm 0.47$ & $2.98 \pm 0.31$ & $2.79 \pm 0.31$ & $-2.73 \pm 0.28$\\ 
\shortgscb &2017-06-01 -- 02:16:29 & $3.07 \pm 0.36$ & $4.62 \pm 0.35$ & $2.95 \pm 0.32$ & $3.32 \pm 0.26$ & $-3.21 \pm 0.28$\\ 
\shortgscb &2017-06-01 -- 02:27:32 & $3.74 \pm 0.36$ & $3.99 \pm 0.38$ & $3.09 \pm 0.30$ & $3.39 \pm 0.31$ & $-2.85 \pm 0.28$\\ 
\shortgscb &2017-06-01 -- 02:44:12 & $3.81 \pm 0.40$ & $4.40 \pm 0.32$ & $3.05 \pm 0.30$ & $3.69 \pm 0.30$ & $-3.60 \pm 0.27$\\ 
\shortgscb &2017-08-19 -- 01:28:50 & $3.16 \pm 0.41$ & $4.76 \pm 0.36$ & $2.92 \pm 0.37$ & $3.08 \pm 0.29$ & $-4.29 \pm 0.29$\\ 
\shortgscb &2017-08-19 -- 01:39:28 & $3.16 \pm 0.43$ & $4.44 \pm 0.41$ & $2.92 \pm 0.40$ & $3.17 \pm 0.32$ & $-4.12 \pm 0.27$\\ 
\shortgscb &2017-08-19 -- 01:55:28 & $2.62 \pm 0.61$ & $5.07 \pm 0.53$ & $3.27 \pm 0.50$ & $2.70 \pm 0.37$ & $-4.14 \pm 0.43$\\ 
\shortgscb &2010-07-09 -- 08:28:56 & - & - & $3.08 \pm 1.05$ & $4.76 \pm 0.92$ & $-13.28 \pm 1.08$\\ 
\shortgscb &2010-07-09 -- 08:34:42 & - & - & $1.18 \pm 0.98$ & $2.30 \pm 0.69$ & $-11.05 \pm 0.72$\\ 
\shortgscb &2010-07-09 -- 08:45:15 & - & - & $3.90 \pm 1.10$ & $3.92 \pm 0.74$ & $-10.36 \pm 0.79$\\ 
\shortgscb &2010-07-09 -- 08:51:02 & - & - & $2.23 \pm 1.20$ & $2.71 \pm 0.85$ & $-9.13 \pm 0.78$\\ 
\shortgscb &2010-07-09 -- 08:59:40 & - & - & $2.14 \pm 0.96$ & $4.04 \pm 0.91$ & $-10.25 \pm 0.79$\\ 
\shortgscb &2010-07-09 -- 09:05:26 & - & - & $3.09 \pm 1.17$ & $3.69 \pm 0.84$ & $-13.89 \pm 0.90$\\ 
\shortgscb &2010-07-09 -- 09:12:50 & - & - & $3.77 \pm 0.80$ & $1.12 \pm 0.82$ & $-9.37 \pm 0.76$\\ 
\shortgscb &2010-07-09 -- 09:18:36 & - & - & $6.39 \pm 1.21$ & $4.93 \pm 1.01$ & $-14.88 \pm 1.17$\\ 
\hline\hline

\end{tabular}
\tablefoot{Measured equivalent widths of the various \KI\ and \PaBeta\ lines, following the method described in \cite{1992ApJS...83..147S} and \cite{2005ApJ...623.1115C}. Date is given in YYYY-MM-DD -- hh:mm:ss format. Possible correlations between the lines are shown in Fig.~\ref{fig:results:KI_correlations}.}
\end{center}
\end{table*}

\begin{table*}
\caption{Gaussian fitting results}             
\label{tab:results:gaussian_fitting}      
\begin{center}
\footnotesize
\begin{tabular}{l c c c c c c}
\hline\hline
Source & Date & I$_0$ & $\upsigma$ & $\upmu_0$ & Int. Flux & Intrinsic FWHM\\ 
& YYYY-MM-DD -- hh:mm:ss &  $10^{-16}$~W/m$^2$/$\upmu$m & km/s & km/s &  $10^{-19}$~W/m$^2$ & km/s\\ 
\hline
\shortgqlupb &2006-09-17 -- 23:22:53 & $25.0 \pm 1.3$ & $63.5 \pm 3.7$ & $-27.8 \pm 3.7$ & $17.0 \pm 0.9$ & $79.1 \pm 16.5$\\ 
\shortgqlupb &2006-09-17 -- 23:39:49 & $19.4 \pm 1.2$ & $79.5 \pm 6.0$ & $-19.0 \pm 5.9$ & $16.5 \pm 1.1$ & $137.5 \pm 19.4$\\ 
\shortgqlupb &2006-09-17 -- 23:45:21 & $20.5 \pm 1.3$ & $72.1 \pm 5.3$ & $-20.4 \pm 5.3$ & $15.8 \pm 1.0$ & $112.6 \pm 18.9$\\ 
\shortgqlupb &2006-09-18 -- 00:02:15 & $22.4 \pm 0.5$ & $67.9 \pm 1.8$ & $-22.5 \pm 1.8$ & $16.3 \pm 0.4$ & $97.0 \pm 7.1$\\ 
\shortgqlupb &2006-09-18 -- 00:07:41 & $19.0 \pm 0.9$ & $75.9 \pm 4.1$ & $-16.8 \pm 4.0$ & $15.4 \pm 0.7$ & $125.6 \pm 13.7$\\ 
\shortgqlupb &2006-09-18 -- 00:24:36 & $25.3 \pm 2.2$ & $54.5 \pm 5.4$ & $-17.1 \pm 5.4$ & $14.8 \pm 1.3$ & $18.8 \pm 87.2$\\ 
\shortgqlupb &2006-09-18 -- 00:30:08 & $22.5 \pm 1.5$ & $65.4 \pm 4.9$ & $-21.2 \pm 4.9$ & $15.8 \pm 1.0$ & $87.2 \pm 20.5$\\ 
\shortgqlupb &2006-09-18 -- 00:47:01 & $29.5 \pm 3.3$ & $61.5 \pm 7.9$ & $-1.3 \pm 7.9$ & $19.4 \pm 2.2$ & $69.3 \pm 39.0$\\ 
\shortgqlupb &2006-09-18 -- 00:52:33 & $45.5 \pm 6.9$ & $30.9 \pm 5.5$ & $-25.6 \pm 5.4$ & $15.1 \pm 2.3$ & $0.0 \pm 58.6$\\ 
\shortgqlupb &2017-04-25 -- 05:21:16 & $5.7 \pm 1.3$ & $38.3 \pm 9.9$ & $-19.2 \pm 9.9$ & $2.3 \pm 0.5$ & $0.0 \pm 80.3$\\ 
\shortgqlupb &2017-04-25 -- 05:26:40 & $4.4 \pm 0.9$ & $63.8 \pm 15.0$ & $-53.8 \pm 14.8$ & $3.0 \pm 0.6$ & $80.4 \pm 66.3$\\ 
\shortgqlupb &2017-04-25 -- 05:37:30 & $4.7 \pm 1.3$ & $46.9 \pm 15.1$ & $-41.4 \pm 15.1$ & $2.4 \pm 0.7$ & $0.0 \pm 101.4$\\ 
\shortgqlupb &2017-04-25 -- 05:42:55 & $4.9 \pm 0.8$ & $43.5 \pm 7.9$ & $-2.6 \pm 7.9$ & $2.3 \pm 0.4$ & $0.0 \pm 71.1$\\ 
\shortgqlupb &2017-04-30 -- 05:34:35 & $4.8 \pm 1.3$ & $47.0 \pm 14.2$ & $-49.0 \pm 14.2$ & $2.4 \pm 0.6$ & $0.0 \pm 98.2$\\ 
\shortgqlupb &2017-04-30 -- 05:56:16 & $3.9 \pm 1.1$ & $45.0 \pm 14.3$ & $-34.3 \pm 14.3$ & $1.9 \pm 0.5$ & $0.0 \pm 98.6$\\ 
\shortgqlupb &2017-04-30 -- 05:50:51 & $5.5 \pm 1.5$ & $40.4 \pm 12.6$ & $-37.1 \pm 12.6$ & $2.4 \pm 0.6$ & $0.0 \pm 91.7$\\ 
\shortgqlupb &2017-04-30 -- 05:40:00 & $4.1 \pm 1.4$ & $35.7 \pm 13.9$ & $-22.4 \pm 13.9$ & $1.6 \pm 0.5$ & $0.0 \pm 96.8$\\ 
\shortgqlupb &2017-06-01 -- 06:29:29 & $9.5 \pm 2.4$ & $49.3 \pm 14.5$ & $-55.9 \pm 14.5$ & $5.0 \pm 1.3$ & $0.0 \pm 99.1$\\ 
\shortgqlupb &2017-06-01 -- 06:34:55 & $7.4 \pm 1.9$ & $52.2 \pm 15.7$ & $-14.2 \pm 15.7$ & $4.1 \pm 1.1$ & $0.0 \pm 103.6$\\ 
\shortgqlupb &2017-06-01 -- 06:45:44 & $11.2 \pm 2.6$ & $56.6 \pm 15.1$ & $-54.0 \pm 15.0$ & $6.8 \pm 1.6$ & $40.1 \pm 118.4$\\ 
\shortgqlupb &2017-06-01 -- 06:51:09 & $11.6 \pm 2.3$ & $48.2 \pm 11.0$ & $-69.2 \pm 10.9$ & $6.0 \pm 1.2$ & $0.0 \pm 85.1$\\ 
\hline 
\shortgscb &2017-04-25 -- 06:21:07 & $3.2 \pm 0.3$ & $54.7 \pm 6.7$ & $-14.2 \pm 6.7$ & $1.9 \pm 0.2$ & $0.0 \pm 74.3$\\ 
\shortgscb &2017-04-25 -- 06:32:12 & $4.0 \pm 0.2$ & $62.4 \pm 2.8$ & $-7.3 \pm 2.7$ & $2.7 \pm 0.1$ & $0.0 \pm 47.2$\\ 
\shortgscb &2017-04-25 -- 06:48:51 & $3.8 \pm 0.3$ & $65.6 \pm 5.5$ & $-10.1 \pm 5.3$ & $2.7 \pm 0.2$ & $0.0 \pm 66.9$\\ 
\shortgscb &2017-04-25 -- 07:35:50 & $3.2 \pm 0.3$ & $62.5 \pm 7.9$ & $-9.7 \pm 7.7$ & $2.1 \pm 0.2$ & $0.0 \pm 80.8$\\ 
\shortgscb &2017-04-25 -- 07:46:53 & $3.2 \pm 0.5$ & $63.7 \pm 11.9$ & $-19.8 \pm 11.6$ & $2.2 \pm 0.3$ & $0.0 \pm 100.8$\\ 
\shortgscb &2017-04-25 -- 08:03:31 & $2.5 \pm 0.2$ & $57.5 \pm 4.5$ & $-11.5 \pm 4.4$ & $1.5 \pm 0.1$ & $0.0 \pm 60.2$\\ 
\shortgscb &2017-05-04 -- 04:29:49 & $3.6 \pm 0.2$ & $74.2 \pm 3.8$ & $-2.4 \pm 3.6$ & $2.9 \pm 0.1$ & $52.9 \pm 29.9$\\ 
\shortgscb &2017-05-04 -- 04:40:54 & $3.7 \pm 0.4$ & $78.6 \pm 10.1$ & $7.7 \pm 9.1$ & $3.1 \pm 0.3$ & $81.0 \pm 54.2$\\ 
\shortgscb &2017-05-04 -- 04:57:38 & $4.5 \pm 0.2$ & $75.7 \pm 3.9$ & $8.2 \pm 3.6$ & $3.7 \pm 0.2$ & $63.5 \pm 25.7$\\ 
\shortgscb &2017-06-01 -- 01:54:27 & $4.4 \pm 0.5$ & $66.4 \pm 8.3$ & $-5.7 \pm 8.1$ & $3.1 \pm 0.3$ & $0.0 \pm 83.3$\\ 
\shortgscb &2017-06-01 -- 02:16:29 & $4.3 \pm 0.3$ & $64.9 \pm 5.1$ & $-0.3 \pm 5.0$ & $3.0 \pm 0.2$ & $0.0 \pm 64.7$\\ 
\shortgscb &2017-06-01 -- 02:27:32 & $4.1 \pm 0.2$ & $72.7 \pm 4.4$ & $5.8 \pm 4.1$ & $3.2 \pm 0.2$ & $40.0 \pm 44.1$\\ 
\shortgscb &2017-06-01 -- 02:44:12 & $4.8 \pm 0.3$ & $78.5 \pm 7.1$ & $13.4 \pm 6.5$ & $4.0 \pm 0.3$ & $80.0 \pm 38.9$\\ 
\shortgscb &2017-08-19 -- 01:28:50 & $6.3 \pm 0.2$ & $73.5 \pm 3.2$ & $-5.9 \pm 3.0$ & $4.9 \pm 0.2$ & $46.9 \pm 28.0$\\ 
\shortgscb &2017-08-19 -- 01:39:28 & $5.8 \pm 0.2$ & $75.3 \pm 4.0$ & $0.3 \pm 3.7$ & $4.7 \pm 0.2$ & $60.8 \pm 27.6$\\ 
\shortgscb &2017-08-19 -- 01:55:28 & $5.5 \pm 0.4$ & $83.5 \pm 8.0$ & $5.7 \pm 7.1$ & $4.9 \pm 0.4$ & $104.4 \pm 35.6$\\ 
\shortgscb &2010-07-09 -- 08:28:56 & $17.4 \pm 0.9$ & $61.1 \pm 3.7$ & $7.5 \pm 3.7$ & $11.4 \pm 0.6$ & $120.2 \pm 10.4$\\ 
\shortgscb &2010-07-09 -- 08:34:42 & $19.6 \pm 1.0$ & $44.5 \pm 2.7$ & $7.2 \pm 2.7$ & $9.4 \pm 0.5$ & $69.1 \pm 9.7$\\ 
\shortgscb &2010-07-09 -- 08:45:15 & $18.1 \pm 1.3$ & $52.6 \pm 4.5$ & $8.1 \pm 4.5$ & $10.2 \pm 0.8$ & $95.4 \pm 13.8$\\ 
\shortgscb &2010-07-09 -- 08:51:02 & $17.2 \pm 0.6$ & $42.0 \pm 1.8$ & $8.5 \pm 1.8$ & $7.8 \pm 0.3$ & $59.8 \pm 6.9$\\ 
\shortgscb &2010-07-09 -- 08:59:40 & $16.9 \pm 0.8$ & $56.6 \pm 3.1$ & $5.3 \pm 3.1$ & $10.2 \pm 0.5$ & $107.4 \pm 9.1$\\ 
\shortgscb &2010-07-09 -- 09:05:26 & $18.1 \pm 1.4$ & $61.1 \pm 5.6$ & $12.0 \pm 5.6$ & $11.8 \pm 0.9$ & $120.2 \pm 15.7$\\ 
\shortgscb &2010-07-09 -- 09:12:50 & $16.0 \pm 0.6$ & $55.2 \pm 2.5$ & $9.8 \pm 2.5$ & $9.4 \pm 0.4$ & $103.2 \pm 7.5$\\ 
\shortgscb &2010-07-09 -- 09:18:36 & $19.6 \pm 1.1$ & $63.7 \pm 4.2$ & $17.9 \pm 4.1$ & $13.4 \pm 0.8$ & $127.7 \pm 11.6$\\ 
\hline\hline

\end{tabular}
\tablefoot{Gaussian fitting results of the \PaBeta\ lines. Date is given in YYYY-MM-DD -- hh:mm:ss format. I$_0$ is the scaling parameter of the gaussian.}
\end{center}
\end{table*}

\end{appendix}

\end{document}